\documentclass[%
 reprint,
superscriptaddress,
nofootinbib,
amsmath,amssymb,
aps,
prx,
longbibliography
]{revtex4-1}

\usepackage{placeins}
\usepackage{appendix}
\usepackage{graphicx}
\usepackage{dcolumn}
\usepackage{bm}
\usepackage{booktabs}
\usepackage{hyperref}
\usepackage{xcolor}
\hypersetup{
    colorlinks,
    linkcolor={blue!70!black},
    citecolor={blue!70!black},
    urlcolor={blue!70!black}
}

\newcommand{\ra}[1]{\renewcommand{\arraystretch}{#1}}

\begin{document}

\title{Molecular Asymmetry and Optical Cycling: Laser Cooling Asymmetric Top Molecules}

\author{Benjamin L. Augenbraun}
\email{augenbraun@g.harvard.edu}
\affiliation{Department of Physics, Harvard University, Cambridge, MA 02138, USA}
\affiliation{Harvard-MIT Center for Ultracold Atoms, Cambridge, MA 02138, USA}

\author{John M. Doyle}
\affiliation{Department of Physics, Harvard University, Cambridge, MA 02138, USA}
\affiliation{Harvard-MIT Center for Ultracold Atoms, Cambridge, MA 02138, USA}

\author{Tanya Zelevinsky}
\affiliation{Department of Physics, Columbia University, New York, NY 10027, USA}

\author{Ivan Kozyryev}
\email{ik2452@columbia.edu}
\affiliation{Department of Physics, Columbia University, New York, NY 10027, USA}

\date{January 29, 2020}

\begin{abstract}
We present a practical roadmap to achieve optical cycling and laser cooling of asymmetric top molecules (ATMs). Our theoretical analysis describes how reduced molecular symmetry, as compared to diatomic and symmetric non-linear molecules, plays a role in photon scattering. We present methods to circumvent limitations on rapid photon cycling in these systems. We calculate vibrational branching ratios for a diverse set of asymmetric top molecules and find that many species within a broad class of molecules can be effectively cooled with a manageable number of lasers. We also describe methods to achieve rotationally closed optical cycles in ATMs. Despite significant structural complexity, laser cooling can be made effective using extensions of the current techniques used for linear molecules. Potential scientific impacts of laser-cooled ATMs span frontiers in controlled chemistry, quantum simulation, and searches for physics beyond the Standard Model.
\end{abstract}

\maketitle

\section{Introduction}

Deviations from perfect symmetry play an important role in Nature at a variety of spatial and temporal scales~\cite{gross1996role}, ranging from the interactions of subatomic particles~\cite{Lee1956,Bouchiat1997} to key biological processes like embryonic laterality~\cite{simunovic20193d,vandenberg2013unified}. Many important biochemical processes in the human body involve chiral molecules, which are necessarily asymmetric tops~\cite{Quack2002, Blackmond2011}. Despite their fundamental importance in other branches of science, asymmetric top molecules (ATMs, molecules with three distinct moments of inertia) have not been put under full quantum control. The traditional tools used to manipulate atomic and molecular samples with exquisite precision generally break down for ATMs due to intrinsic structural complexity.

Highly symmetric, linear molecules have been brought under quantum control via recent advances producing laser-cooled molecules in the $\mu$K regime. This has spurred progress in such diverse fields as ultracold chemistry~\cite{krems2010viewpoint,krems2008cold,balakrishnan2016perspective}, quantum simulation of strongly correlated systems~\cite{Carr2009}, and precision tests of fundamental physics~\cite{demille2015diatomic,kozyryev2017PolyEDM}. In many cases, the rich variety of structures afforded by many molecules (e.g., large, permanent body-frame electric dipole moments and closely-spaced pairs of levels with opposite parity) conveys fundamental advantages over atomic alternatives. Direct laser cooling of molecules is a new, important tool for reaching the ultracold regime and achieving full control over the quantum state of the molecule. Furthermore, cooling allows for the long interaction times required to exploit these structures.\footnote{The technique of optoelectric Sisyphus cooling~\cite{prehn2016} is a complementary method to cool polyatomic molecules. However, it relies on neither directional momentum exchange between laser fields and molecules nor cycling optical photons, and is therefore orthogonal to the ideas described in this work.}  SrF~\cite{Barry2014,norrgard2015sub}, CaF~\cite{Truppe2017b,truppe2017CaF, Anderegg2017CaFMOT, Cheuk2018lambda}, and YO~\cite{Collopy2018} have been directly laser-cooled, trapped, and had their quantum states detected with near unit efficiency. The linear triatomic molecules SrOH~\cite{kozyryev2016radiation, kozyryev2016Sisyphus}, YbOH~\cite{AugenbraunYbOHSisyphus}, and CaOH~\cite{Baum2020} have been laser cooled in one dimension, the first step to full quantum control.

Direct laser cooling requires a closed optical cycling transition such that $> 10^4$ photons may be scattered using a reasonable number of lasers and without significant loss to dark manifolds of states~\cite{Tarbutt2018}. To date, only linear molecules have been directly laser cooled. The high degree of symmetry in these molecules greatly restricts the decay channels that could remove a molecule from the optical cycle. Extending the technique to \textit{asymmetric} molecules necessarily involves removing these restrictions. While Isaev and Berger previously performed \textit{ab initio} estimates of the vibrational branching ratios for the chiral ATM MgCHDT~\cite{isaev2015polyatomic}, to date there has been no complete description of general methods to directly laser cool and trap ATMs. It has not previously been clear how the reduced symmetry of ATMs would affect optical cycling, or even if full laser cooling of asymmetric species would be possible.

Although potentially challenging to produce, ultracold ATMs would offer many qualitatively unique features useful for a broad range of science. For example, ATMs can have up to three permanent dipole moments that could conceivably be controlled independently, a feature never present in higher symmetry species. Low-lying states with very long radiative lifetimes and dipole moments $\sim$5~D have been observed, promising strong molecule-molecule coupling~\cite{Scurlock1994}. In addition, asymmetric molecular structures often arise when various chemically interesting ligands (e.g., NH$_2$, SH, OC$_2$H$_5$, OCH(CH$_3$)$_2$, C$_4$H$_5$N, etc.) are bonded to the optical cycling center atom. Finally, chiral molecules are, by necessity, ATMs. The determination of a general method to laser cool ATMs--- allowing a ligand to be chosen primarily for its targetted utility rather than technological accessibility--- could open up new avenues in fields including quantum information/simulation, ultracold chemistry, and precision measurements (see Sec.~\ref{sec:Applications}).

In this paper, we demonstrate that a general class of ATMs is amenable to laser cooling and trapping techniques. We focus on species of the form \hbox{M-L}, comprising an alkaline-earth atom (M) that is ionically and monovalently bonded to an electronegative ligand (L). In this case the metal-centered unpaired valence electron is expected to act somewhat independently of the bond, leading to good optical cycling properties~\cite{kozyryev2016MOR,Ivanov2019Rational}. We present calculations of the vibrational branching ratios in these molecules, demonstrating that with only a few repumping lasers one may scatter enough photons to slow, trap, and cool the molecules to the Doppler limit. Once at that temperature, these molecules will be generically amenable to sub-Doppler techniques, allowing access to the $\mu$K regime. Crucially, we describe how, despite the complex rotational structure in these asymmetric tops, a simple laser cooling scheme may be constructed that limits rotational branching to an easily repumped manifold of states. Using available spectroscopic data and new \textit{ab initio} quantum chemistry calculations we discover that, surprisingly, in many cases the required experimental complexity is not significantly greater than that encountered with linear polyatomic molecules. Finally, we elaborate on some applications of trapped, ultracold ATMs that our work opens.

\begin{figure}[tb]
\centering 
\includegraphics[width=0.8\columnwidth]{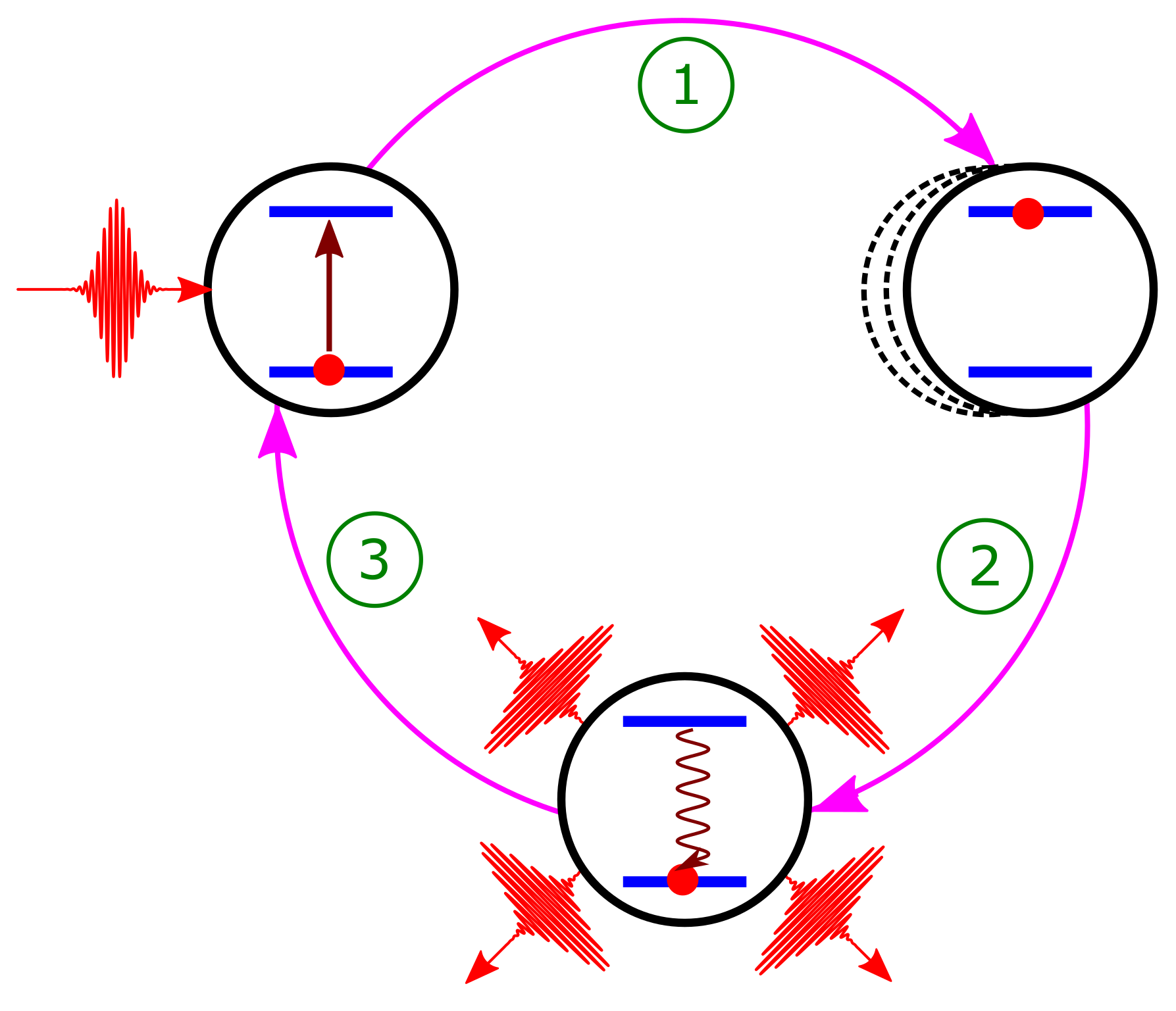}
\caption{Schematic diagram of optical cycling in an ideal two-level system. Repetition of steps 1-3 are required to achieve radiative laser cooling: 1. directional photon absorption, 2. isotropic spontaneous emission, and 3. decay to the initial quantum level. The complex structure of molecules can interrupt this cycle, e.g. by the addition of multiple decay pathways during step 3.}
\label{fig:PhotonCycling}
\end{figure}

\section{Achieving Photon Cycling}
Photon cycling, the repeated absorption and emission of photons, is a fundamental necessity to laser cooling (see Fig.~\ref{fig:PhotonCycling}). In this section we discuss how the achievement of photon cycling is complicated by molecular structure. We present here techniques to overcome these complications. 

Working within the Born-Oppenheimer approximation, we separate the discussion of electronic, vibrational, and rotational excitations. The possibility of Born-Oppenheimer approximation breakdown and its challenges are discussed in App.~\ref{sec:BOBreakdownApp}.

Laser cooling relies on photon cycling to generate dissipative forces~\cite{McCarron2018,Tarbutt2018}. Typically $\sim10^4 - 10^5$ photon scatters are necessary to apply sufficient force for slowing, cooling, and trapping. This requires an excited state that decays preferentially to one ground state, with unplugged leaks to all other (possibly metastable) levels below about 1 part in $10^4$. In the case of atoms, the simplest with this electronic structure are alkali or alkaline-earth (AE) atoms with one or two valence electrons in $s$ orbitals. Electronic branching is then limited, or wholly eliminated, so that repumping from ``dark" levels is easily achieved with at most a few additional laser frequencies. In the case of molecules, certain classes with $^2 \Sigma$ ground states have been found to be favorable to laser cooling due to their single valence electron in a metal-centered $s\sigma$ orbital~\cite{McCarron2018, Tarbutt2018}. This leads to a simple electronic structure that closely resembles that of alkali metal atoms.

Vibrational and rotational substructure within each electronic state can be significant impediments to photon cycling in molecules. These internal degrees of freedom can change quantum number during an electronic transition, leading to the possibility of quick diffusion of population into ``dark states," i.e. states that are not coupled to the applied laser light. Molecules lost to these states will not be cooled further unless they are repumped, a task usually accomplished by application of additional laser or microwave frequencies. For linear molecules, angular momentum selection rules can eliminate losses to dark rotational levels~\cite{Stuhl2008}. Losses to dark vibrational states are, by contrast, always problematic due to the general lack of selection rules. Vibrational branching during an electronic decay is governed probabilistically by Franck-Condon factors (FCFs), $q_{v^\prime,v^{\prime\prime}}$, which characterize the overlap between excited and ground vibrational states.\footnote{Excited states are denoted by a single prime while ground states are denoted by double primes.} Vibrational losses are highly suppressed in molecules for which the ground and excited state potential energy surfaces resemble one another, as they do for AE metal atoms monovalently and ionically bonded to an electronegative ligand (M$^+$L$^-$). In these species, electrostatic repulsion of the M$^{+}$-centered electron by the ligand anion induces orbital hybridization which moves the electron away from the chemical bond and strongly decouples electronic and vibrational excitations~\cite{Ellis2001,Bernath1991, Bernath1997}.

These ideas have led to the successful laser cooling of many AE-monofluorides and AE-monohydroxides (all linear), including SrF~\cite{Shuman2009,Shuman2010, Barry2012, Barry2014, McCarron2015, norrgard2015sub, Steinecker2016, McCarron2018}, CaF~\cite{Truppe2017b, truppe2017CaF, Anderegg2017CaFMOT,  Williams2017, Cheuk2018lambda, Caldwell2019}, YbF~\cite{Lim2018}, SrOH~\cite{kozyryev2016radiation,kozyryev2016Sisyphus}, YbOH~\cite{AugenbraunYbOHSisyphus}, and CaOH~\cite{Baum2020}. Isaev and Berger proposed laser cooling symmetric top molecules including CaCH$_3$ and MgCH$_3$ in 2016~\cite{isaev2015polyatomic}, considering only the vibrational branching ratios. Recently, FCF calculations were reported for species with AE atoms bonded to hydrocarbon chains or fullerenes, showing that these species have vibrational branching ratios favorable for laser cooling~\cite{Klos2019}. Ref.~\cite{kozyryev2016MOR} considered the issues of rotational and fine structure in symmetric, nonlinear molecules. It was found that laser cooling symmetric top molecules requires at most one additional (rotational) repumper as compared to diatomic species. For ATMs it is not obvious \textit{a priori} that typical laser cooling techniques will be applicable because, as the degree of symmetry of a molecule is decreased, selection rules become progressively weaker; many possible rotational- and vibrational-state changing decays are allowed during an electronic transition. Any discussion of photon cycling in ATMs therefore requires careful consideration of both vibrational and rotational loss channels.

In ATMs, all three principal axes have distinct moments of inertia. These molecules can have at most twofold rotational symmetry and, formally, there can be no orbitally degenerate states in these species.\footnote{This includes point groups $C_1$, $C_i$, $C_s$, $C_2$, $D_2$, $C_{2h}$, $D_{2h}$, and $C_{2v}$.} The reduced symmetry has several important consequences for laser cooling. First, mixing between electronic manifolds--- especially among the closely-spaced electronically excited states $\tilde{A}$, $\tilde{B}$, and $\tilde{C}$--- may lead to perturbations that need to be understood and controlled in order to scatter a large number of photons. Second, vibronic selection rules that are present in the case of linear or symmetric top molecules may begin to break down. Third, the reduced symmetry requires careful consideration when attempting to construct a rotationally closed cycling transition because angular momentum selection rules are relaxed. We analyze potential loss channels related to electronic, vibrational, and rotational structure separately in the following discussion.

\begin{figure}[tb]
\centering 
\includegraphics[width=\columnwidth]{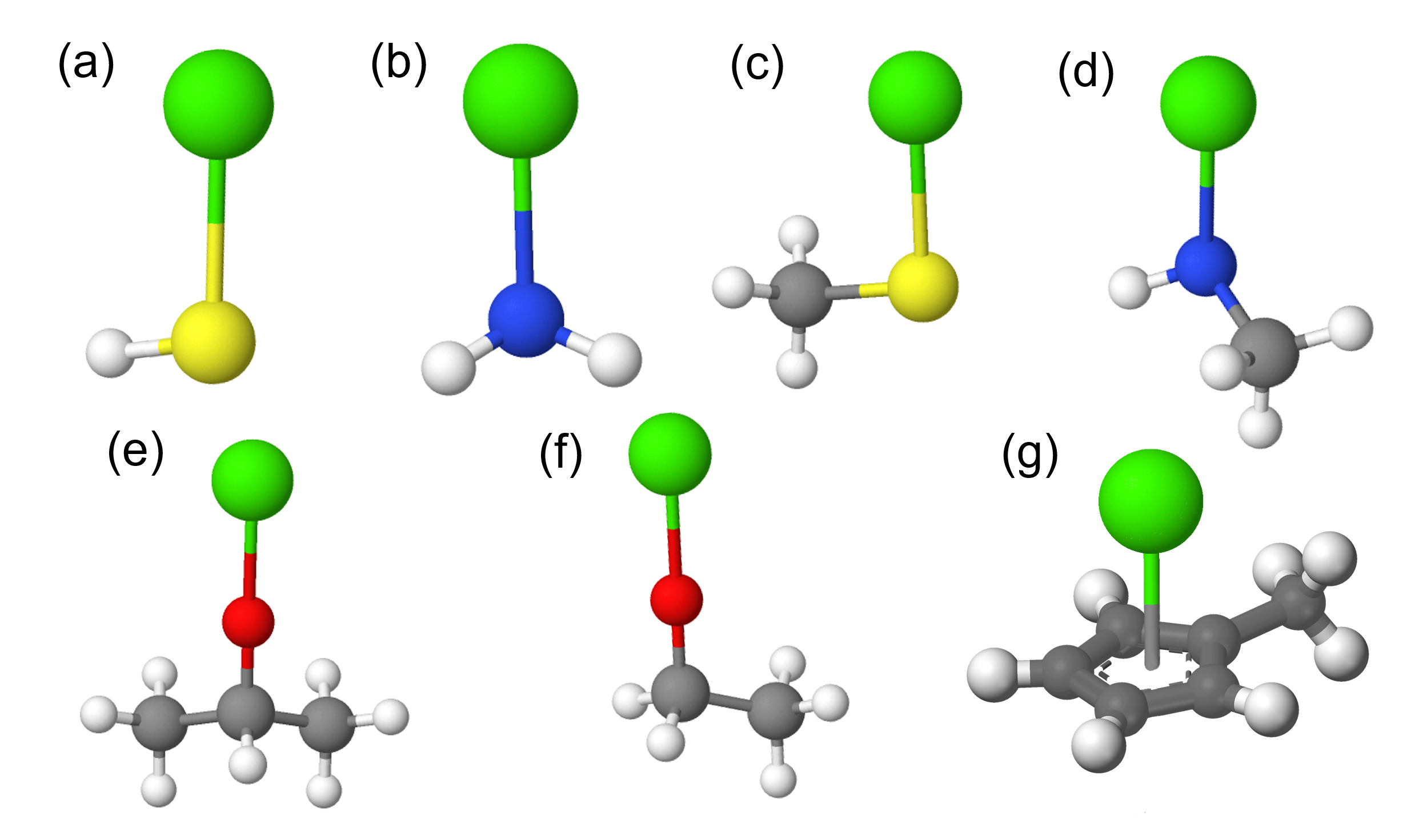}
\caption{Model geometries for several of the species proposed in this work. (a) MSH, (b) MNH$_2$, (c) MSCH$_3$, (d) MNHCH$_3$,  (e) MOCH(CH$_3$)$_2$, (f) MOC$_2$H$_5$, (g) MC$_5$H$_4$CH$_3$, where M represents an alkaline-earth metal atom. Atomic species are colored according to green: alkaline-earth metal, yellow: sulfur, blue: nitrogen, red: oxygen, gray: carbon, white: hydrogen. Structures were generated using MolView~\cite{MolView}.}
\label{fig:MoleculeShapes}
\end{figure}

We consider here a very large class of molecules which has fortunately been the focus of considerable previous spectroscopic attention (making practical application in the laboratory much easier). Such species include (see Fig.~\ref{fig:MoleculeShapes}): AE monoamides (MNH$_2$)~\cite{Marr1995,Morbi1998,Brazier2000,Morbi1997,Sheridan2005SrNH2, Sheridan2000,Sheridan2001, Wormsbecher1983}, hydrosulfides (MSH)~\cite{Jarman1993, Sheridan2007, Shirley1990, TalebBendiab2001, Janczyk2003}, monoethoxides (MOC$_2$H$_5$)~\cite{Paul2016}, pyrollyls (MC$_4$H$_4$N)~\cite{Bopegedera1990}, methylpentadienyls (MC$_5$H$_4$CH$_3$)~\cite{Robles1992,Bopegedera1990,Cerny1995}, methanethiols (MSCH$_3$)~\cite{Fernando1991}, alkylamides (MNHCH$_3$)~\cite{Bopegedera1987alkylamide}, and isopropoxides (MOCH(CH$_3$)$_2$)~\cite{brazier1986laser}, where M = Mg, Ca, Sr, or Ba.\footnote{Ca and Sr-containing species have received the most attention. Some Mg and Ba-containing species have been observed, e.g. MgSH~\cite{TalebBendiab2001}, BaSH~\cite{Janczyk2003}, MgNH$_2$~\cite{Sheridan2000,Sheridan2001}, and BaNH$_2$~\cite{Wormsbecher1983}. Yb-containing ATMs are interesting targets due to the high-$Z$ nucleus, but have not yet been studied.} Chiral-substituted AE-methyls (MCHDT)~\cite{Sheridan2005CaCH3,Salzberg1999,Rubino1995} and AE-methoxides (MOCHDT)~\cite{OBrien1988} are also considered because the chiral methyl group will cause the three moments of inertia to differ. The wealth of spectroscopic data available for these species is essential to providing quantitative examples for the general approach to laser cooling ATMs that we describe here. 

\subsection{Electronic Transitions} \label{sec:Electronic}

\begin{figure}[t!]
\centering
\includegraphics[width=\columnwidth]{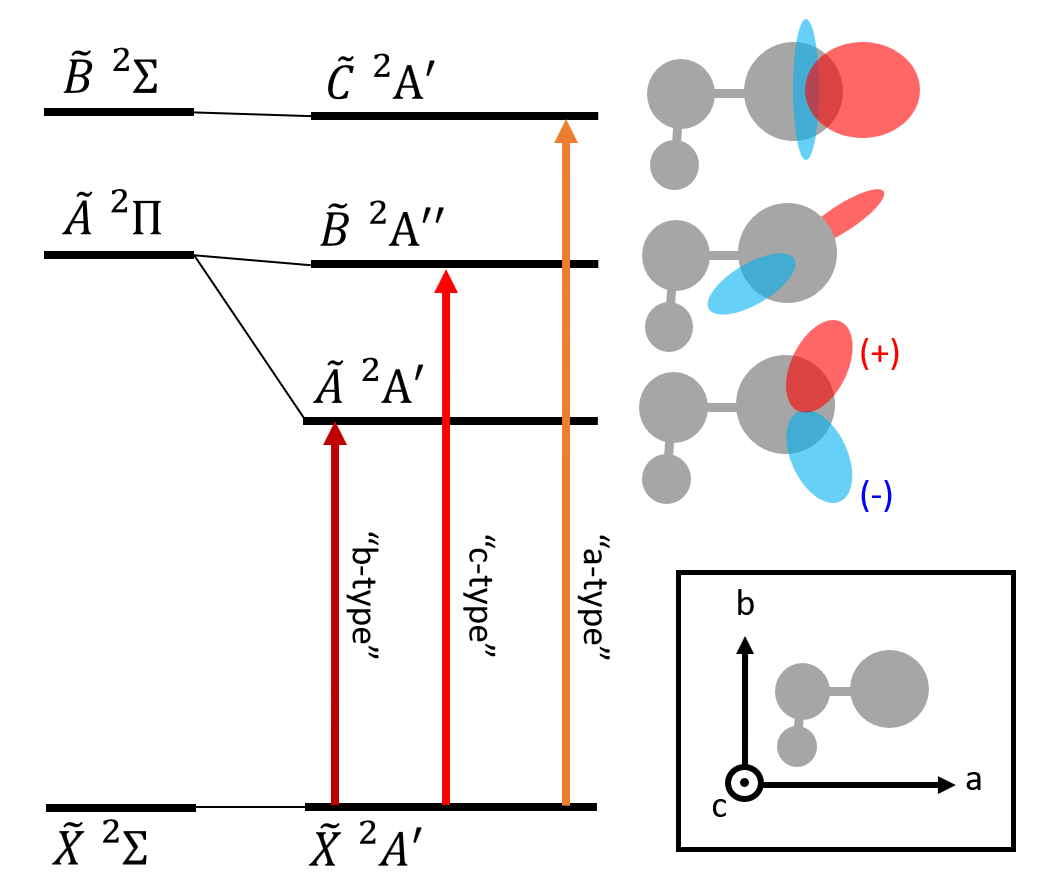}
\caption{Energy level diagram for species of $C_s$ symmetry relevant to the present work, including correlations between linear (left) and asymmetric top (right) states. The asymmetry of the molecule leads to a splitting of the $^2\,\Pi$ potential, lifting the degeneracy of the in-plane and out-of-plane orbitals. The diagram is essentially the same, aside from symmetry labeling, for other highly ionic doublet species.  Optical transitions are labeled by the dominant transition dipole moment component. Schematic drawings of the orientation of the electron density are included to rationalize these assignments. Red and blue label phase of the electronic wavefunction. Inset: Principal axis system for the example species. 
}
\label{fig:ElectronicStates}
\end{figure}

For the molecules considered in this work, the structure of the low-lying electronic states can be understood through their correlations with the electronic states of linear AE monohalides~\cite{Ellis2001,Morbi1997}. A schematic energy level diagram is shown in Fig.~\ref{fig:ElectronicStates}. Aside from the exact symmetry species labeling, the same pattern of levels will hold for any of the ATMs considered here. As the  AE-metal atom, M, approaches the ligand, L, one of the $ns^2$ valence electrons is transferred from M to L. The electron remaining localized around M$^{+}$ is polarized away from the bond via orbital mixing with excited $np$ and $nd$ orbitals~\cite{Ellis2001}. In the linear limit, the ground state, $\tilde{X}$, is of $^2\Sigma^+$ symmetry while the two lowest electronically excited states are $\tilde{A}\,^2\Pi$ and $\tilde{B}\,^2\Sigma^+$. As the cylindrical symmetry of the molecule is broken, the degeneracy between orbitals directed along the $b$- and $c$-axes is lifted and the $\tilde{A}\,^2\Pi$ state splits in two. 

In general there will be four low-lying states of interest to laser cooling experiments, $\tilde{X}\,^2A^\prime$, $\tilde{A}\,^2A^{\prime}$, $\tilde{B}\,^2A^{\prime\prime}$, and $\tilde{C}\,^2A^\prime$ in the case of $C_{s}$ symmetry. (The symmetry labeling will differ depending on the point group, but the overall structure will be similar.) For molecules of $C_{2v}$ symmetry, like MNH$_2$, the principal axes of the molecule coincide with the axes along which the electronic orbitals are aligned. For molecules of lower symmetry, there will be a misalignment between the M-L bond and the primary axis along which the electronic orbitals are oriented. As will be discussed below, this angle determines to what extent the rotational selection rules expected within a given electronic transition will hold. For the species considered here, the lowest three electronically excited states can be addressed by convenient laser wavelengths ranging between 570~nm and 730~nm and have short lifetimes $\sim20-40$~ns. The visible wavelengths and large spontaneous decay rates are crucial for exerting large optical forces.

\begin{figure}[t!]
\centering
\includegraphics[width=\columnwidth]{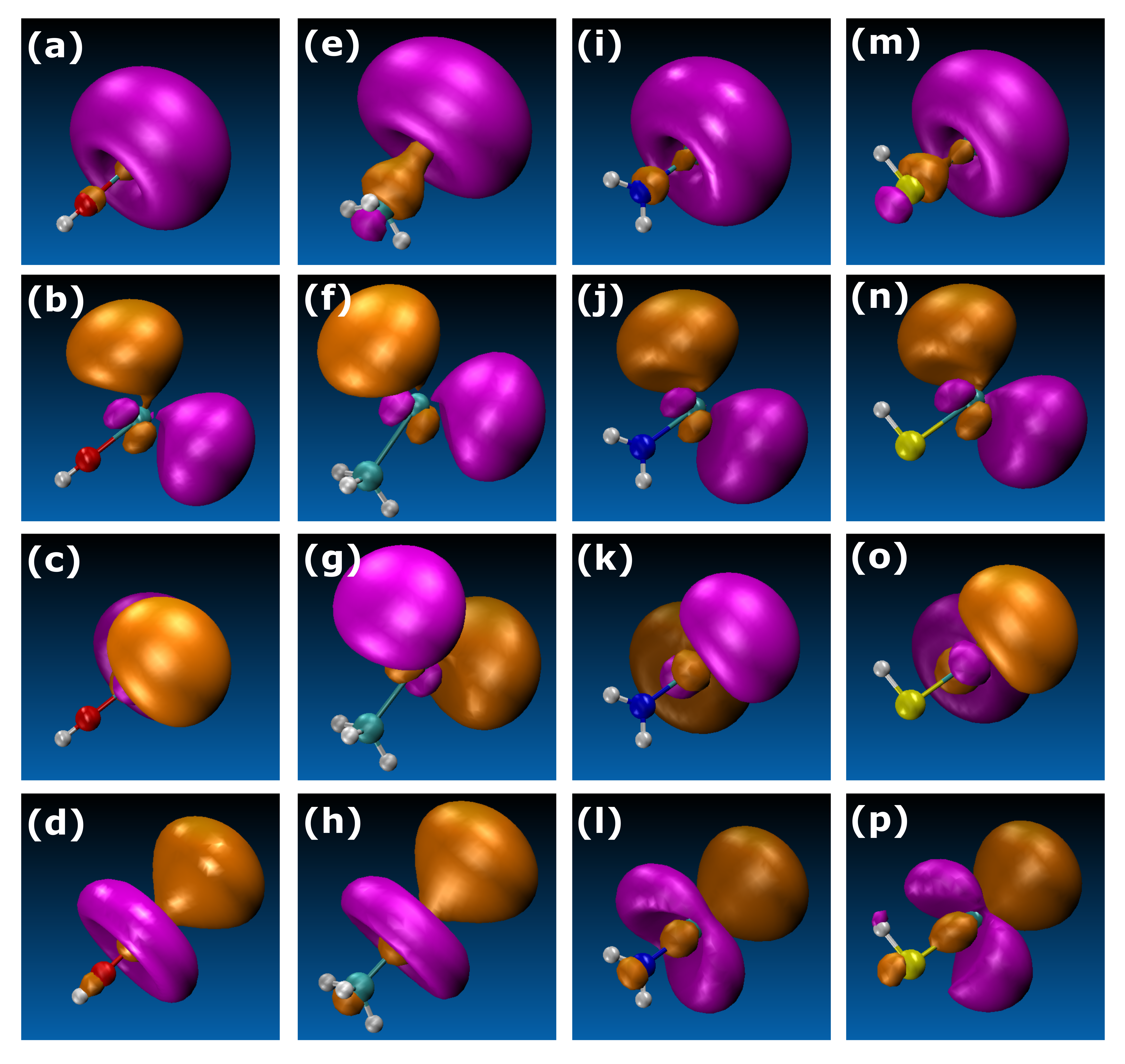}
\caption{Comparison of molecular orbitals for electronic transitions employed in the proposed laser cooling scheme for four ML molecules with different symmetries: CaOH (a-d), CaCH$_3$ (e-h), CaNH$_2$ (i-l), and CaSH (m-p). Molecular geometries are plotted for the ground electronic state; rows 1 through 4 correspond to  SOMO, LUMO, LUMO+1, and LUMO+2 configurations, respectively. Molecular symmetry is incrementally reduced from column 1 to column 4: $C_{\infty v}$, $C_{3v}$, $C_{2v}$, and $C_s$. All isosurfaces are plotted for the isovalue of 0.03 with magenta (orange) representing positive (negative) values. 
}
\label{fig:MolOrbs}
\end{figure}

In order to understand the detailed electronic structure of laser-coolable ATMs, we have performed \textit{ab initio} molecular structural calculations for four prototypical species: CaOH ($C_{\infty v}$ symmetry), CaCH$_3$ ($C_{3v}$ symmetry), CaNH$_2$ ($C_{2v}$ symmetry) and CaSH ($C_s$ symmetry). These are the four simplest species within each symmetry group considered. Figure~\ref{fig:MolOrbs} provides a comparison of the singly occupied molecular orbital (SOMO) and the lowest unoccupied molecular orbital (LUMO) as the symmetry of the ML molecule is reduced (column $1\rightarrow4$). Despite significant differences in structural symmetry, for the SOMO, LUMO and LUMO+1 (rows 1-3) the molecular orbitals are essentially unchanged as the molecule distorts. The orbitals are only slightly distorted for LUMO+2 (row 4). Quantitatively, both ionic bonding and heavy valence electron localization on the metal atom are described by the Mulliken charge and spin population analyses. As the ligand is changed from OH ($C_{\infty v}$) to CH$_3$ ($C_{3v}$) to NH$_2$ ($C_{2v}$) and SH ($C_s$), the Mulliken charge on the Ca atom goes from $+0.58$ to $+0.58$ to $+0.55$ and $+0.51$ while the spin population changes from $0.999$ to $0.908$ to $0.982$ and $0.963$, respectively. This is in qualitative agreement with the electron affinities determined for these anion ligands~\cite{Vamhindi2016}. The similar electronic structure in these molecules suggests that laser cooling of ATMs will proceed similarly to cooling linear species. 

Losses due to purely electronic transitions are not expected to affect photon cycling in these molecules. One potential loss channel involves any electronic state intermediate to the ground and excited states used for optical cycling. Radiative decay through these intermediate states will result in a change of parity and populate dark rovibrational states in $\tilde{X}$~\cite{Collopy2018}. For the molecules considered here, no electronic states between the $\tilde{A}$ and $\tilde{X}$ states have been observed. When optical cycling with the excited $\tilde{B}$ or $\tilde{C}$ states, one may still worry about radiative cascades through the lower-lying $\tilde{A}$ state. Due to the small energy spacing between the three lowest excited states ($100 - 1000$~cm$^{-1}$), radiative decays rates from $\tilde{B}$ or $\tilde{C}$ to $\tilde{A}$ will be suppressed by a factor of $\left( \Delta \omega / \omega \right)^{-3} \sim 10^4$ relative to decays directly to $\tilde{X}$. Thus, loss due to potential two-photon cascades will come in at the same level as that observed already in experiments cooling diatomic species~\cite{Truppe2017b}, which does not present any challenges to cooling and trapping experiments. 

A second potential source of loss involves higher-order multipole transitions, e.g. M1 and E2 transitions, that would remove molecules from the optical cycle by populating opposite parity rovibrational states. M1 transitions cannot change the electronic configuration and are therefore $\sim \omega_{M1}^3 \alpha^{2} / \omega_{E1}^3 \sim 10^{-9}$ smaller than E1 transitions rates, where $\alpha \approx 1/137$ is the fine-structure constant~\cite{Chae2015}. The rate of E2 transitions is expected to be $\sim (\omega/c)^2 \langle f \rvert r^2 \lvert i \rangle / \langle f \rvert r \lvert i \rangle \sim 10^{-7}$ that of E1 transitions, where $\lvert i \rangle$ and $\lvert f \rangle$ are the initial and final states, respectively. Therefore, loss due to M1 or E2 transitions should be negligible until $\gg 10^6$ photons have been scattered. Such effects have not yet limited diatomic laser cooling experiments, and are unlikely to affect laser cooling of ATMs~\cite{Tarbutt2018,McCarron2018}.

\subsection{Vibrational Branching} \label{sec:Vibrational}
Accurate values of the vibrational branching ratios during an electronic decay are of utmost important to molecular laser cooling. To date, very few measurements of vibrational branching ratios for ATMs have been performed. To provide information relevant to laser cooling, these measurements would ideally use cw excitation to a low-$J$ rotational state, be free from state-changing collisions, and include careful calibration of detection efficiency over a wide range of wavelengths. These restrictions limit the number of useful direct FCF measurements in the literature. In contrast (and fortunately), high-quality spectroscopic data (including bond lengths and angles) for many molecules of interest is plentiful. As such, we use a semi-empirical method to compute FCFs for a large set of ATMs expected to be favorable for laser cooling. When accurate high-resolution spectroscopic measurements are available, the computational methods described below can provide vibrational branching ratios in excellent agreement with observations, and with accuracy better than purely \textit{ab initio} predictions~\cite{kozyryevCaOCH3, nguyen2018fluorescence}. The accuracy of these calculations depends entirely on the accuracy of the measured molecular geometries. In the absence of such experimental data, \textit{ab initio} calculations have enabled unique insights into identification of favorable ligands for laser cooling~\cite{Ivanov2019Rational} and estimation of the vibrational branching ratios~\cite{Isaev2017},

Franck-Condon factors, which characterize the probability of vibrational state changes in an electronic decay, are computed by evaluating the overlap integral between excited and ground vibrational wavefunctions:

\begin{equation}
    q_{v^\prime,v^{\prime\prime}} = \left \lvert \int \psi_{v^\prime}(\mathbf{Q}^\prime) \psi_{v^{\prime\prime}}(\mathbf{Q}^{\prime\prime}) \, d\mathbf{Q}^{\prime\prime} \, d\mathbf{Q}^\prime \right \rvert^2, 
    \label{eq:FCIntegral}
\end{equation}

\noindent where $\mathbf{Q}^{\prime\prime}$ represent the vibrational normal modes of the ground state and $\mathbf{Q}^{\prime}$ the vibrational normal modes of the excited state. We compute these integrals within the harmonic approximation, which is justified given that anharmonic contributions are $\omega_e x_e / \omega_e \approx 5 \times 10^{-2}$ for the AE-containing molecules considered here are ~\cite{Fernando1991}.

\begin{figure}[t!]
\centering 
\includegraphics[width=\columnwidth]{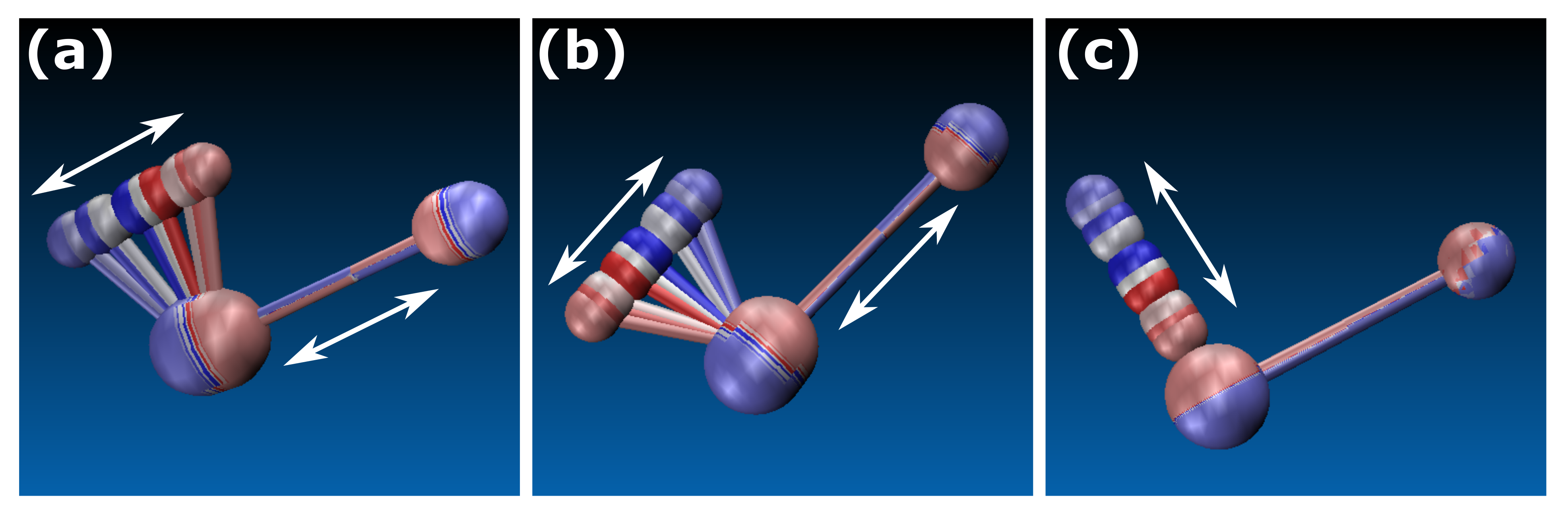}
\caption{Normal modes for CaSH corresponding to (a) 316~cm$^{-1}$, (b) 360~cm$^{-1}$ and (c) 2640~cm$^{-1}$ vibrational frequencies. The beginning of the trajectory is indicated in red, the middle in white, and the end in blue. Note the strong mixing between the bending motion and Ca-S stretching vibration.}
\label{fig:CaSHmodes}
\end{figure}

We follow standard methods to calculate the FCFs~\cite{Chen1994}. In brief, we first perform a $\mathbf{GF}$-matrix analysis of the vibrational problem~\cite{WilsonDeciusCross} to determine the normal modes in each electronic manifold. Experimentally measured vibrational frequencies are used to fit the elements of the $\mathbf{F}$ matrix. We include a Duschinsky transformation between ground and excited normal coordinates, which accounts for normal mode mixing (see Fig.~\ref{fig:CaSHmodes}). We then follow the method of Sharp and Rosenstock~\cite{Sharp1964,OberlanderThesis} which uses generating functions and a linear transformation between the normal coordinates of ground and excited states to define analytic expressions for the FCFs. Analytical formulas evaluating Eq.~\ref{eq:FCIntegral} are available for various combinations of vibrational excitations~\cite{Sharp1964,Weber2003}. This model allows one to predict vibrational branching ratios for all vibrational modes within the harmonic approximation. It has been validated for many linear and non-linear polyatomic species~\cite{kozyryevCaOCH3, AugenbraunYbOHSisyphus, nguyen2018fluorescence} with excellent agreement between theory and experiment. We find that the computed FCFs depend strongly on the changes in molecular geometry upon electronic excitation and more weakly on changes in vibrational frequencies. High-resolution spectroscopy confirms that the molecular geometries change minimally upon electronic excitation~\cite{Marr1995,Brazier2000,Morbi1997}. 

\begin{table}[t]
\ra{1.3}
\begin{tabular*}{\columnwidth}{@{}l@{\extracolsep{\fill}}lcc@{}} 
\hline \hline Transition                        & Decay to & \multicolumn{1}{c}{Caculated}            & \multicolumn{1}{c}{Measured\footnote{From Ref.~\cite{Brazier2000}}} \\ \hline 
$\tilde{A} \rightarrow \tilde{X}$ & $0_0$ &  0.960 & 0.959\\
                                  & 1x Sr-N stretch    & 0.039  & 0.04 \\
                                  & 2x Sr-N stretch & $1 \times 10^{-3}$ & $\sim 1.6 \times 10^{-3}$ \\
                                  & 1x NH$_2$ bend            & $3 \times 10^{-5}$ & $2 \times 10^{-5}$      \\
                                  & 1x N-H sym. stretch         & $8 \times 10^{-6}$ & $4 \times 10^{-5}$       \\ \hline
$\tilde{C} \rightarrow \tilde{X}$ & $0_0$ & 0.976 & - \\
                                  & 1x Sr-N stretch   & 0.011  & 0.01  \\
                                  & 1x N-H sym. stretch      & $6 \times 10^{-6}$   & $8 \times 10^{-6}$       \\ 
                                  & 1x NH$_2$ bend & 0.010 & overlapped  \\ 
                                  \hline \hline 
\end{tabular*}
\caption{Comparison of measured and calculated vibrational branching ratios for SrNH$_2$. Because decay to the NH$_2$ bending mode is overlapped in the $\tilde{C} \rightarrow \tilde{X}$ band, we are not able to sum experimental FCFs in this case. The molecular geometry and vibrational frequencies used in the calculations are taken from Refs.~\cite{Brazier2000, Thompsen2000, Sheridan2005SrNH2}.}
\label{tab:SrNH2FCF}
\end{table}

We benchmark our multidimensional FCF calculations by comparing to experimental measurements. FCFs for SrNH$_2$ have been previously measured~\cite{Brazier2000} on the $\tilde{C} \, ^2A_1 \rightarrow \tilde{X} \, ^2A_1$ and $\tilde{A} \, ^2B_2 \rightarrow \tilde{X} \, ^2A_1$ bands. Table~\ref{tab:SrNH2FCF} reports the calculated vibrational branching ratios for SrNH$_2$ and their comparison to experimentally measured values; we find excellent agreement between theory and experiment. We speculate that the small discrepancies, which occur in vibrations of the ligand, arise from the fact that the spectroscopy was typically unable to fully determine the ligand geometry, so systematic errors affecting the calculations are possible. These calculations indicate that for SrNH$_2$, photon cycling of $\sim 10^4$ photons should be possible with just three vibrational repumping lasers. There also exist measurements of vibrational branching ratios in CaOC$_2$H$_5$~\cite{Paul2016}, predicting a decay to the Ca-O stretching mode of $\sim 0.1$, in good agreement with our calculation below.\footnote{These measurements were taken with pulsed-laser excitation and the data shows signs of collisional excitation which could present systematic errors in their interpretation.} Besides these few cases, vibrational branching ratios for nonlinear, laser-coolable species are rarely reported. Clearly, there is need for new, accurate measurements of vibrational branching ratios in AE-containing ATMs. 

Calculated multidimensional FCFs for CaNH$_2$, CaSH, and SrSH are reported in Tab.~\ref{tab:GFModelFCF}. Although there are no published FCF measurements for these species, high-resolution studies have noted that off-diagonal decays were either unobservably small or highly suppressed, in agreement with our predictions~\cite{Fernando1991}. For these molecules, our calculations indicate that 3-5 repumping lasers will be required to scatter $>10^4$ photons. Generally, the calculations agree very well with the simpler model used for the predictions in Tab.~\ref{tab:MLFCF}. Where differences occur, these can be attributed to mixing of normal modes that was neglected in the simple model~\cite{Nicholls1981}. Vibrational normal modes of MSH species are shown in Fig.~\ref{fig:CaSHmodes}, depicting a strong mixing between the bending and M-S stretching motions; this mixing is included in our calculations. Dynamic visualizations of all vibrational normal modes for CaNH$_2$ are provides in the media Supplemental Materials. 
Proper accounting of the differences between ground state and excited state normal modes is included in the calculations of Tabs.~\ref{tab:SrNH2FCF}~and~\ref{tab:GFModelFCF}. Comparison of the methods also shows that while the individual FCFs may differ slightly, the sum of the dominant few FCFs is consistent.

\begin{table}[t]
\ra{1.3}
\begin{tabular*}{\columnwidth}{@{}l@{\extracolsep{\fill}}ccc@{}} \hline \hline
& \multicolumn{3}{c}{CaNH$_2$}                                                                                                                                             \\ \cline{2-4} 
                & $\tilde{A} \rightarrow \tilde{X} $    & $\tilde{B} \rightarrow \tilde{X} $    & $\tilde{C} \rightarrow \tilde{X} $                     \\ \cline{1-4} 
$0^0_0$         & 0.963  & 0.993    & 0.979                                                   \\
1x Ca-N stretch & 0.034   & $2 \times 10^{-3}$    & 0.019                                                   \\
1x NH$_2$ bend & $2 \times 10^{-3}$ & $1 \times 10^{-3}$ & $8.8 \times 10^{-4}$ \\
2x Ca-N stretch & $1.6 \times 10^{-4}$ & $5 \times 10^{-5}$ & $3.6 \times 10^{-6}$\\ 
1x N-H sym. stretch & $3.1 \times 10^{-7}$ & $3 \times 10^{-6}$ & $3.3 \times 10^{-7}$ \\ \cline{1-4} 
                
                & \multicolumn{3}{c}{CaSH}                                                                                                                                                 \\ \cline{2-4} 
                & \multicolumn{1}{c}{$\tilde{A} \rightarrow \tilde{X} $} & \multicolumn{1}{c}{$\tilde{B} \rightarrow \tilde{X} $} & \multicolumn{1}{c}{$\tilde{C} \rightarrow \tilde{X} $} \\ \cline{1-4} 
$0^0_0$         & 0.820                                                  & 0.952                                                  & 0.999                                                  \\
1x Ca-S stretch & 0.163                                                  & 0.0276                                                 & $3.3 \times 10^{-4}$                                     \\
1x Ca-S-H bend  & 0.016                                                  & 0.0199                                                 & $3.5 \times 10^{-6}$                                   \\
2x Ca-S stretch & $1.6 \times 10^{-4}$                                   & $4\times 10^{-4}$                                      & $< 10^{-6}$     \\ 
1x S-H stretch & $1 \times 10^{-6}$ & $1 \times 10^{-6}$ & $1 \times 10^{-6}$ \\ \hline
                & \multicolumn{3}{c}{SrSH}                                                                                                                                                 \\ \cline{2-4} 
                & \multicolumn{1}{c}{$\tilde{A} \rightarrow \tilde{X} $} & \multicolumn{1}{c}{$\tilde{B} \rightarrow \tilde{X} $} & \multicolumn{1}{c}{$\tilde{C} \rightarrow \tilde{X} $} \\ \cline{1-4} 
$0^0_0$         & 0.828                                                 & 0.953                                                  & 0.976                                                  \\
1x Sr-S stretch & 0.170                                                  & 0.046                                                & 0.023                                     \\
1x Sr-S-H bend  & 0.0014                                                 & $4 \times 10^{-4}$                                 & $2 \times 10^{-4}$                             \\
2x Sr-S stretch & $1 \times 10^{-4}$                                                  & $<10^{-5}$                                     & $< 10 ^{-4}$     \\
1x S-H stretch & $1 \times 10^{-6}$ & $1 \times 10^{-6}$ & $1 \times 10^{-6}$ \\
\hline \hline 

\end{tabular*}
\caption{Calculated multidimensional FCFs for CaNH$_2$, CaSH, and SrSH. The molecular geometry and vibrational frequencies used in the calculations are as reported in Refs.~\cite{Sheridan2007, Jarman1993,Fernando1991,Shirley1990, Morbi1997, Morbi1998,Marr1995}.}
\label{tab:GFModelFCF}
\end{table}

The computations above require the complete molecular geometry and vibrational frequencies for all vibrational modes, which have not been experimentally determined for many species. Often, only the metal-ligand bond lengths have been determined. Experience with the  monohydroxides shows that the metal-ligand stretching mode is typically the dominant decay channel. Table~\ref{tab:MLFCF} presents metal-ligand stretching mode FCFs for a wide variety of species~\cite{Nicholls1981}. The FCFs for these ATMs are highly diagonal due to the small changes in potential energy surface shapes upon decay. We estimate the relative error in our calculations by using the same model for compute FCFs for isoelectronic diatomic and linear triatomic species; for the diagonal FCFs, the relative error is expected to be $< 3\%$. Note that perturbations in the excited electronic state can introduce unexpectedly strong off-diagonal decays~\cite{kozyryevCaOCH3}, and for any particular species the dominant uncertainty in the FCFs will likely be due to these perturbative couplings. These will become more prevalent especially with increasing complexity and decreasing symmetry of the ligand (see App.~\ref{sec:BOBreakdownApp}).

\begin{table}[t]
\ra{1.3}
\begin{tabular*}{\columnwidth}{@{}l@{\extracolsep{\fill}}ccccc@{}} 
\hline \hline Molecule           & Transition                       & $q_{00}$ & $q_{01}$ & $\lambda_{00}$ (nm) & $\lambda_{01}$ (nm)  \\ \hline
CaSH        & $\tilde{A} \rightarrow \tilde{X}$ & 0.826     & 0.157     & 650                 & 664                 \\
            & $\tilde{B} \rightarrow \tilde{X}$ & 0.983     & 0.016     & 630                 & 643                 \\
            & $\tilde{C} \rightarrow \tilde{X}$ & 0.999     & 0.0003    & 622                 & 634                 \\ \hline
SrSH        & $\tilde{A} \rightarrow \tilde{X}$ & 0.850     & 0.137     & 700                 & 713                 \\
            & $\tilde{B} \rightarrow \tilde{X}$ & 0.962         & 0.037         & 675                 & 687                 \\
            & $\tilde{C} \rightarrow \tilde{X}$ & 0.981     & 0.0184    & 666                 & 678                 \\ \hline
MgNH$_2$\footnote{Mg-N bond lengths inferred from Refs.~\cite{Wright1999, Sheridan2000, Sheridan2001}.}    & $\tilde{A} \rightarrow \tilde{X}$ & 0.933     & 0.065     & $\sim$410\footnote[2]{Excitation wavelengths not measured at high resolution.}                 & $\sim$420\footnotemark[2]               \\ \hline 
CaNH$_2$    & $\tilde{A} \rightarrow \tilde{X}$ & 0.964     & 0.035     & 640                 & 662                 \\
            & $\tilde{B} \rightarrow \tilde{X}$ & 0.997     & 0.0002    & 633                 & 654                 \\
            & $\tilde{C} \rightarrow \tilde{X}$ & 0.976     & 0.022     & 576                 & 593                 \\ \hline
SrNH$_2$    & $\tilde{A} \rightarrow \tilde{X}$ & 0.957         & 0.039         & 700                 & 723                 \\
            & $\tilde{B} \rightarrow \tilde{X}$ & 0.971     & 0.028     & 679                 & 700                 \\
            & $\tilde{C} \rightarrow \tilde{X}$ & 0.956     & 0.013     & 630                 & 648                 \\ \hline
BaNH$_2$\footnote[3]{Ba-N bond lengths inferred from Refs.~\cite{Wormsbecher1983, Ivanov2019Rational}.}    & $\tilde{A} \rightarrow \tilde{X}$ & 0.916         & 0.081         & $\sim$895\footnotemark[2]                 & $\sim$925\footnotemark[2]                  \\
            & $\tilde{C} \rightarrow \tilde{X}$ & 0.851     & 0.137     & $\sim$760\footnotemark[2]                  & $\sim$780\footnotemark[2]                  \\  \hline
MgCHDT      & $\tilde{A} \rightarrow \tilde{X}$ & 0.936     & 0.062    & 499                 & 526    \\ \hline
CaCHDT      & $\tilde{A} \rightarrow \tilde{X}$ & 0.997     & 0.002    & 680                 & 700  \\\hline
CaOCHDT   & $\tilde{A} \rightarrow \tilde{X}$ & 0.951     & 0.048    & 628                 & 648                 \\
                   & $\tilde{B} \rightarrow \tilde{X}$ & 0.946    & 0.046     & 565                 & 586    \\ \hline
SrOCHDT  & $\tilde{A} \rightarrow \tilde{X}$ & 0.945     & 0.046    & 689                 & 709                 \\ \hline
CaSCH$_3$  & $\tilde{A} \rightarrow \tilde{X}$ & 0.809    & 0.171    & 645                 & 658                 \\
                   & $\tilde{B} \rightarrow \tilde{X}$ & 0.981   & 0.018    & 633                 & 645    \\ \hline
SrSCH$_3$  & $\tilde{A} \rightarrow \tilde{X}$ & 0.832    & 0.153    & 694                 & 706                 \\
           & $\tilde{B} \rightarrow \tilde{X}$ & 0.978   & 0.021    & 676                 & 688    \\ 
           & $\tilde{C} \rightarrow \tilde{X}$ & 0.957   & 0.042    & 647                 & 657    \\\hline
CaNHCH$_3$  & $\tilde{A} \rightarrow \tilde{X}$ & 0.953    & 0.046    & 652                 & 673  \\ \hline
SrNHCH$_3$  & $\tilde{A} \rightarrow \tilde{X}$ & 0.944    & 0.054    & 706                 & 726   \\ \hline
CaOC$_2$H$_5$ & $\tilde{A} \rightarrow \tilde{X}$   & 0.892         & 0.102         & 631                  & 647  \\ \hline
CaOCH(CH$_3$)$_2$  & $\tilde{A} \rightarrow \tilde{X}$ & 0.956    & 0.043    & 632                 & 645 \\ \hline
SrOCH(CH$_3$)$_2$  & $\tilde{A} \rightarrow \tilde{X}$ & 0.943    & 0.056    & 690                 & 704 \\ \hline
CaC$_5$H$_4$CH$_3$\footnote[4]{Dispersed fluorescence measured in Ref.~\cite{Robles1992}.}  & $\tilde{A} \rightarrow \tilde{X}$ & 0.825    & 0.094    & 691                 & 706 \\
& $\tilde{B} \rightarrow \tilde{X}$ & 0.86   & 0.078    & 686                 & 701 \\
\hline \hline 

\end{tabular*}
\caption{Franck-Condon factors calculated for decay to 0 quanta ($q_{00}$) and 1 quantum ($q_{01}$) of the metal-ligand stretching mode. $\lambda_{00}$ and $\lambda_{01}$ are the main cycling and first repumping transition wavelengths in nm. Experimentally measured geometries and frequencies come from Refs.~\cite{Marr1995,Morbi1998,Brazier2000,Morbi1997,Sheridan2005SrNH2, Sheridan2000,Sheridan2001, Wormsbecher1983, Jarman1993, Sheridan2007, Shirley1990, TalebBendiab2001, Janczyk2003, Paul2016, Bopegedera1990,Robles1992,Bopegedera1990,Cerny1995,Fernando1991,Bopegedera1987alkylamide,TalebBendiab2001,Janczyk2003,Sheridan2000,Sheridan2001,Wormsbecher1983,Salzberg1999,Rubino1995}.}
\label{tab:MLFCF}
\end{table}

We have also conducted a set of \textit{ab initio} calculations to understand these species (see App.~\ref{sec:AbInitioApp}). Normal modes and vibrational frequencies were computed at optimized geometries using the ORCA Quantum Chemistry Software~\cite{neese2012orca} and wavefunction overlap intergrals were computed numerically using the ezSpectrum software~\cite{mozhayskiy2009ezspectrum}. The calculations are benchmarked by testing the predictions for CaOH against experimentally determined FCFs~\cite{kozyryevCaOCH3}. For the lowest excited electronic transition $\tilde{A}\rightarrow \tilde{X}$, the FCF for the fundamental $0^0_0$ transition changed from 0.988 (CaOH) to 0.870 (CaSH) to 0.989 (CaNH$_2$). The dominant vibrational loss channel was to the Ca-ligand stretching mode for all three molecules with decay fractions 0.01 (CaOH), 0.10 (CaSH) and 0.01 (CaNH$_2$) to this mode. The Duschinsky transformation~\cite{Kupka1986} of the normal modes was used for all three molecules to account for the change in normal coordinates in each electronic state. In the case of CaSH, we observed strong mixing between the stretching and bending motions (see Fig.~\ref{fig:CaSHmodes}). While the CaSH $0^0_0$ vibronic FCF is predicted to be $<0.9$, the sum of the three dominant vibrational loss channels is $\gtrsim0.99$. Previous work~\cite{Isaev2017} has indicated that only the sum of the three dominant FCFs should be considered stable for predictive purposes when purely \textit{ab initio} methods are being used.

Based on these calculations and their agreement with experiment, ATMs are strong candidates for optical cycling and laser cooling. For species of the form MSH and MNH$_2$ (M=Mg, Ca, Sr), the calculations indicate that at least one electronic transition in several of the species considered can scatter $>10^3$ photons with only a single repumping laser. The calculations predict that the number of photon scatters per molecule is increased to $>10^4$ per molecule--- the typical number required to slow a molecular beam and capture into a MOT--- with three repumping lasers. Indeed, the calculated FCFs are comparable to those of isoelectronic diatomic species which have already been successfully laser-cooled~\cite{Truppe2017b,norrgard2015sub}. For the larger species, at least 100 photons can be scattered with just two lasers. It is remarkable that even in species as complex as calcium methylcyclopentadienyl [CaC$_5$H$_4$CH$_3$, Fig.~\ref{fig:MoleculeShapes}(g)], with 36 vibrational normal modes, the FCFs converge to $\gtrsim 0.95$ with just one or two repumping lasers~\cite{Robles1992}. This is sufficient for high-fidelity detection, preparation of single quantum states, or transverse laser cooling in order to enhance interrogation times in a molecular beam~\cite{kozyryev2016Sisyphus,Lim2018} (see Sec.~\ref{sec:PhotonBudgets}).

\subsection{Rotational Branching} \label{sec:Rotational}
Laser cooling requires, in addition a vibrationally closed optical cycle, rotationally closed transitions. The rotational structure of ATMs can be extremely complex (see App.~\ref{sec:RotationalApp}). Following standard convention, we label the rotational states by $N_{K_a K_c}$. $N$ is the total angular momentum excluding electron and nuclear spins. The principal rotational constants are labeled $A$, $B$, and $C$, where $A > B > C$. $K_a$ and $K_c$ are not rigorous quantum numbers, but are defined as the projection of $N$ onto the hypothetical symmetry axis formed by distorting the molecule to prolate ($a$-axis) and oblate limits ($c$-axis), respectively. Note that a Hund's case (b) basis is often most convenient as $N$ is an approximately good quantum number. Due to the unpaired electron spin, which can couple weakly to rotation, $J=N+S$ is the total angular momentum excluding nuclear spin. Hyperfine structure is typically unresolved ($<1$~MHz). Losses to dark states throughout the complicated rotational ladder are a potential concern during photon cycling. Here we show that rotational and parity selection rules can eliminate or mitigate these potential problems.

\begin{table}[b]
\begin{tabular*}{\columnwidth}{@{}l@{\extracolsep{\fill}}ccc@{}} 
\hline \hline
         & $\Delta K_a$ & $\Delta K_c$ & Exceptions   \\ \hline
$a$-type & 0            & $\pm$ 1      & $\Delta N \neq 0$ for $K_a^\prime \rightarrow K_a^{\prime\prime}=0$ \\
$b$-type & $\pm$ 1      & $\pm$ 1      &                                                                     \\
$c$-type & $\pm$ 1      & 0        &  $\Delta N \neq 0$ for $K_c^\prime \rightarrow K_c^{\prime\prime}=0$ \\ \hline \hline
\end{tabular*}
\caption{Selection rules in the dipole approximation for transitions proposed ATM laser-cooling transitions. Higher order decays ($\Delta K_c = \pm 2$, etc.) are suppressed, but may be induced by perturbations. There is an additional restriction due to the fact that $K_a+K_c = N$ or $N+1$~\cite{Cross1944, HerzbergVol3}.}
\label{tab:RotSelectionRules}
\end{table}

In a generic ATM, the transition dipole moment, $\hat{\mu}$, can have components along any of the three principal axes. The transition dipole moment components are labeled $\hat{\mu}_a$, $\hat{\mu}_b$, and $\hat{\mu}_c$ to indicate their projection along the principal axes of the molecule. The transitions are then labeled $a$-type, $b$-type, or $c$-type according to which component of the transition dipole moment they couple. For a molecule nearer the prolate limit, $b$- and $c$-type transitions are ``perpendicular" while $a$-type transitions are ``parallel." The selection rules for each type of transition are summarized in Tab.~\ref{tab:RotSelectionRules}. There are additional restrictions due to the fact that $K_a+K_c=N$ or $N+1$~\cite{Cross1944,HerzbergVol3}. The selection rules and their symmetry considerations are discussed in detail in App.~\ref{sec:RotationalBranchingApp}.

\begin{figure}[t]
\centering
\includegraphics[width=1\columnwidth]{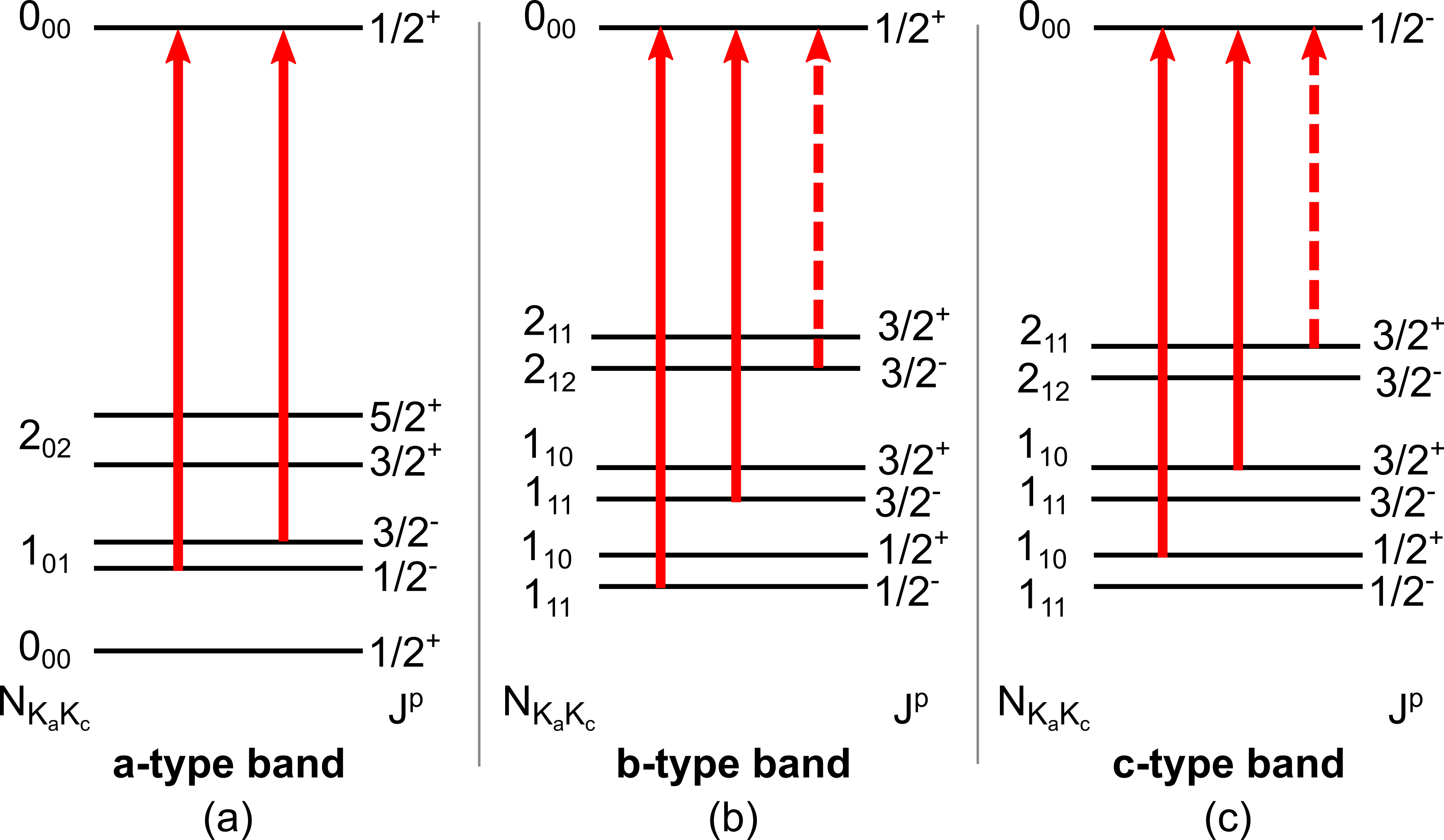}
\caption{Schemes to achieve rotationally-closed photon cycling for pure (a) $a$-type transitions, (b) $b$-type transitions, and (c) $c$-type transitions. Dashed lines indicate transitions which will need to be repumped in special cases, as described in the text. In the $K_a=1$ levels, the spin-rotation and asymmetry doubling are approximately the same size. These level diagrams assume a near-prolate ATM, but analogous results apply for near-oblate tops.} 
\label{fig:RotationalRepumping}
\end{figure}

Using the selection rules enumerated in Tab.~\ref{tab:RotSelectionRules} we find that it is possible to construct closed cycling schemes for transitions coupling to any component of $\hat{\mu}$. Some representative photon cycling schemes are presented in Fig.~\ref{fig:RotationalRepumping}. Dashed lines indicate rotational branches which may be induced by spin-rotation mixings in some cases (see App.~\ref{sec:RotationalBranchingApp} for intensity estimates). In well-behaved cases, the dashed lines may not require repumping. An $a$-type band, expected for the $\tilde{C} \leftarrow \tilde{X}$ transitions, is shown in Fig.~\ref{fig:RotationalRepumping}(a). The rotational structure is reminiscent of a $(b)^2\Sigma^+ \leftarrow \, (b)^2\Sigma^+$ transition in linear molecules, where the first letter indicates the appropriate Hund's case.\footnote{Such a transition is used for laser slowing of CaF molecules~\cite{Truppe2017b} and laser cooling of SrOH molecules~\cite{kozyryev2016Sisyphus}.} In this case, a combination of parity and angular momentum selection rules guarantee rotational closure when driving the $^{Q}Q_{12}(0,1,0.5)$ and $^{Q}P_{11}(0,1,1.5)$ lines.\footnote{We use the branch designation $^{\Delta K_a} \Delta J_{F_n^{\prime} F_n^{\prime \prime}} \left(K_a^{\prime \prime}, K_c^{\prime \prime}, J^{\prime \prime} \right)$, where $F_n$ labels the spin-rotation component.} A single rf-sideband, easily generated using an electro-optic modulator (EOM) or acousto-optic modulator (AOM), must be added to the excitation laser. Figure~\ref{fig:RotationalRepumping}(b) shows a scheme for rotational closure on a $b$-type band, typical of $\tilde{A} \leftarrow \tilde{X}$ transitions. Here, excitation out of the $K_a^{\prime \prime}=1$ sublevel is convenient because it allows one to target the lowest level of the excited electronic state. In this case, we drive the $^{P}Q_{12}(1,1,0.5)$ and $^{P}P_{11}(1,1,1.5)$ lines. Figure~\ref{fig:RotationalRepumping}(c) shows how to achieve rotational closure on a $c$-type band, applicable to the $\tilde{B} \leftarrow \tilde{X}$ transitions. The $b$- and $c$-type bands shown here are similar to a $(b)^2\Sigma^{\pm} \leftarrow \, (b)^2 \Pi$ transition in linear species.\footnote{Such a transition is proposed for laser cooling the CH radical~\cite{Wells2011}.} Here, we choose to drive the $^{P}Q_{12}(1,0,0.5)$ and $^{P}P_{11}(1,0,1.5)$ lines. Similar to the $a$-type band, an rf sideband is required to address the two spin-rotation components of the $N^{\prime \prime}=1$ manifold for either $b$-type or $c$-type bands. Note that the $K_a^{\prime \prime} = 1$ state, which is convenient for optical cycling, provides a suitable ground state because it is metastable with lifetimes $\gg 10$~s for the species considered here.

The cycling schemes in Fig.~\ref{fig:RotationalRepumping} provide a basis for rotational closure in realistic molecules. Roughly speaking, $\tilde{A}\rightarrow \tilde{X}$ decays follow $b$-type selection rules, $\tilde{B} \rightarrow \tilde{X}$ decays obey $c$-type selection rules, and $\tilde{C} \rightarrow \tilde{X}$ decays obey $a$-type selection rules. As the molecule becomes more asymmetric these patterns will break down and a given excited state may decay via one of several components of $\hat{\mu}$. A hybrid $ab$-type band will commonly occur for molecules with $C_s$ symmetry due to the misalignment of the principal axes from the electronic orbitals [see Fig.~\ref{fig:MolOrbs}(m-p)]. One additional decay must be repumped, in essence combining the schemes in Fig.~\ref{fig:RotationalRepumping}(a) and (b). Alternatively, microwave remixing within the ground manifold can return population to the optical cycle. The degree of asymmetry will determine the relative strengths of different components of $\hat{\mu}$. For example, because $\hat{\mu}_a \gg \hat{\mu_c}$ when photon cycling on the $\tilde{C} \leftarrow \tilde{X}$ band of hydrosulfide species, these can scatter $\sim 10^4$ photons using the scheme of Fig.~\ref{fig:RotationalRepumping}(a) before repumping of the $K_a^{\prime \prime}=1$ state is required (see Apps.~\ref{sec:AbInitioApp} and \ref{sec:RotationalBranchingApp}). For hybrid $bc$-type decays, which may occur in molecules with symmetry lower than $C_s$ or due to perturbations, repumping is easily achieved by combining the schemes of Fig.~\ref{fig:RotationalRepumping}(b) and (c). A single laser with sidebands imposed by EOMs or AOMs can bridge the full spin-rotation/asymmetry doubling structure. Finally, for species with near-$C_{3v}$ symmetry (e.g. MOCHDT or MCHDT), where the $\tilde{A} \, ^2A^{\prime}$ and $\tilde{B} \, ^2 A^{\prime \prime}$ states have essentially coalesced into the $\tilde{A} \, ^2E$ level, an additional decay to the $2_{1 K_c} (J^{\prime \prime}=3/2)$ state will appear. Again, this state can be remixed using either microwaves or a single additional laser frequency, as shown by dashed lines in Fig.~\ref{fig:RotationalRepumping}. The effects of perturbations, though not expected to significantly alter the results presented above, are discussed in App.~\ref{sec:RotationalBranchingApp}. Overall we see that, despite the complex ATM rotational structure, application of one or two lasers (depending on the molecular species) with appropriate rf sidebands leads to rotational closure sufficient for scattering $\gtrsim 10^6$ photons. Remarkably, this level of complexity is comparable to that required for laser cooling of diatomic species for which rotational repumping is implemented~\cite{norrgard2015sub, Steinecker2016, Yeo2015, Collopy2018}.

\subsection{General Principles}
The detailed calculations of the previous section focused on select species that have been observed in prior experiments. These calculations elucidate many general principles which apply to the whole class of ATMs with optical cycling centers. 

First, the key features of the electronic structure in the species discussed here derive from the metal-centered valence electron, with potential energy surfaces which vary little from one electronic state to another. Our calculations show that the ligand's asymmetry does not strongly perturb the electronic distribution; the electronic distribution is similar to that expected from a simple diatomic model, independent of the structural asymmetry of the ligand. This crucial feature leads to the possibility of photon cycling. Molecules containing other AE or AE-like atoms, e.g. Be, Mg, Ba, or Yb, should have analogous electronic structure and be favorable for laser cooling. Certain of these AE (or AE-like) atoms may be ideal for particular applications (see Sec.~\ref{sec:Applications}). For instance, the low mass of Be could be beneficial to studies involving torsional modes, while the high-$Z$ Yb nucleus can enhance symmetry violating effects. The relative insensitivity of the electronic structure to the AE atom's identity means that the ideas presented above apply to a very wide range of species, aiding the task of searching for promising candidates. Beyond moleules with AE atoms, several laser-coolable molecules have been identified containing cycling centers from group 3 or 13 of the periodic table, notably TiO~\cite{Stuhl2008},  AlF~\cite{Truppe2019}, TlF~\cite{Hunter2012}, and BH~\cite{Hendricks2014}. We expect that ATM species isoelectronic to these (AlSH, TlOCH$_2$CH$_3$, etc.) will also have highly diagonal FCFs. However, to date such species have received little or no spectroscopic attention. 

Second, although the FCFs do not follow selection rules present for linear species, we have shown that they are still highly constrained and amenable to photon cycling/ laser cooling. In general, the off-diagonal decay strengths do not increase as the molecule becomes ``more" asymmetric. Rather, the strengths remain weak so long as the potential energy surfaces are similar in ground and excited state. Based on the geometry changes measured by high-resolution spectroscopy and predicted by our \textit{ab initio} calculations, this is the case for the broad class of molecules explored here. As expected, it is the \textit{change} in geometry which drives the FCFs, rather than the geometry itself~\cite{Sharp1964, Chen1994}. While a dedicated vibrational analysis will be required for any particular molecule, our calculations indicate that dozens of ATM species are highly amenable to photon cycling. While vibrational anharmonicity is small for the molecules considered here, its presence may be an important consideration if ``floppier" ATMs are of interest in the future. This will particularly be the case for species which include relatively long carbon chains.

Third, we have seen how the leakages due to rotational branching can be mitigated or wholly eliminated by working with low-$J$ states and approximate rotational selection rules. While we considered primarily the near-prolate limit, molecules near the oblate limit behave analogously but with $K_a$ and $K_c$ interchanged. Even for chiral molecules near the prolate (or oblate) limit, the same rotational closure schemes are expected to be applicable. The strength of hybrid rotational branches is expected to increase with the degree of structural asymmetry. Additionally, off-diagonal terms in the rotational Hamiltonian (see App.~\ref{sec:RotationalApp}) may admix levels with $\Delta K_a = \pm 2$ and $\Delta K_c = \pm 2$ selection rules. However, these matrix elements conserve $J$ so that for the low-$J$ states considered here the Hamiltonian is still highly restricted. Transitions with $\Delta K_a = \pm 2, \pm 3, \cdots$ (and similarly for $\Delta K_c$) are limited by the restriction that $K_a + K_c = N$ or $N+1$. In the absence of strong perturbations, even highly asymmetric species will likely require at most one or two additional laser or microwave frequencies to achieve rotational closure. The rotational level structure for any given molecule will be determined by the particular values of $A$, $B$, and $C$, and the levels will not necessarily group into nearly-spaced sets of levels labeled by $K_a$ (prolate) or $K_c$ (oblate). As shown above, in the absence of strong perturbations our proposed rotational pumping schemes require an additional frequency (either microwave or optical) per vibrational band. 

\section{Experimental Outlook}

\subsection{Molecule production} \label{sec:Production}
The molecules considered above can be generated in a straightforward manner using the cryogenic buffer-gas cooling method, which has been used in all molecular laser cooling experiments to date~\cite{hutzler2012buffer, Tarbutt2018, McCarron2018}. Ablation of a metallic target into a reactant gas, followed by buffer-gas cooling due to collisions with cryogenic He can be used, as has been demonstrated with Ca + SF$_6$~\cite{Truppe2017b} and Yb + CH$_3$OH~\cite{AugenbraunYbOHSisyphus}. Commercially available reactants such as hydrogen sulfide, ammonia, chiral methyl fluoride, thiols, pentadienyls, etc. can be introduced with a heated capillary to produce the species considered above. Typical yields of $\sim 10^{10}$ molecules per ablation pulse have been demonstrated for a variety of molecules~\cite{hutzler2012buffer}. Production can be enhanced by an order of magnitude by exciting the ablated metal atoms to the metastable $^3P_1$ state~\cite{Fernando1991, Bopegedera1987alkylamide, Jadbabaie2019}. Especially for larger molecules, buffer-gas cooling provides invaluable phase-space compression and limits the number of populated vibrational modes such that significant molecular population is available for manipulation.

This internal cooling enhances the population in low-lying rotational states by several orders of magnitude without significantly boosting the forward velocity of a beam extracted from the cryogenic cell. Rotational cooling is an important consideration when deciding on the exact photon cycling scheme to employ. For species without reflection symmetry, population will be cooled into the lowest $K_a^{\prime\prime} = 0$ state, with a smaller amount of population in $K_a^{\prime \prime} = 1$. On the other hand, molecules with $C_{2v}$ symmetry (e.g., MNH$_2$) will cool into both the $K_a^{\prime\prime}=0$ or $K_a^{\prime\prime}=1$ levels due to nuclear spin statistics as previously observed in supersonic beams~\cite{Sheridan2005SrNH2}. Because populating the $K_a^{\prime \prime}=1$ level is often useful for photon cycling, we emphasize that it is straightforward to populate this level for any of the species discussed here using microwaves, optical pumping, or natural population. 

\subsection{From Photon Cycling to Laser Cooling} \label{sec:PhotonBudgets}
For the molecules considered in Tab.~\ref{tab:MLFCF} (and assuming a reasonable number of lasers), the number of photons that can be scattered before loss to a dark state can vary considerably. It is therefore worthwhile to enumerate the advantages of optical cycling over a range of different ``photon budgets."

First, consider the case where vibrational leakage channels are closed at the $10^{-2}$ level; this allows $\sim10-100$ photon scatters per molecule before significant loss to a vibrational dark state. This level of photon scattering allows quantum state preparation and high-fidelity readout, allowing 100\% detection fielity for each molecule in a beam~\cite{DeMille2019, lasner2018statistical}. Well-developed optical pumping and coherent population transfer techniques~\cite{panda2016stimulated} would be enabled, a significant advantage to molecular beam experiments described below. Given the FCFs calculated above, each of the molecules considered in Tab.~\ref{tab:MLFCF} would fall in this regime given an experiment using a single vibrational repumping laser.

If vibrational losses are controlled at the $10^{-3}$ level, $\sim 1,000$ photons may be scattered by each molecule before significant loss to a dark state. Such a photon budget is sufficient for transverse laser cooling of linear molecules, decreasing molecular beam divergence and increasing phase-space density by at least an order of magnitude~\cite{AugenbraunYbOHSisyphus, kozyryev2016Sisyphus, Lim2018}. Feasible coherence times for measurements using transversely cooled molecules would increase by factors of $10-100$. In addition, this level of photon scattering would make possible slowing of a molecular beam to the capture velocity of a molecular MOT using novel techniques like the coherent bichromatic force or pulsed laser deceleration~\cite{kozyryev2017BCF,Long2019,galica2018deflection} or a Zeeman-Sisyphus decelerator~\cite{Fitch2016}. For the molecules proposed in Tab.~\ref{tab:MLFCF}, scattering $\sim10^3$ photons will likely require 3-5 lasers, strongly dependent on the particular molecule.

A ``full" laser cooling experiment can be modeled after recent work in diatomic species~\cite{Tarbutt2018,McCarron2018}. Given typical molecular masses ($\sim 50-100$~amu) and buffer-gas beam source velocities ($\sim 70-100$~m/s), roughly $20,000$ photon scatters are required to slow a molecule to rest. Furthermore, at typical scattering rates of $\sim10^6$~s$^{-1}$, another $5,000-10,000$ photons are required for capture into a magneto-optical trap and cooling to the Doppler limit. Modeling loss to vibrational dark states as a Poisson process, in order to scatter $\sim 30,000$ photons without losing more than about half of the molecular population requires repumping all decays that appear at the $>5\times10^{-5}$ level. Such a level of photon scattering would enable the full range of techniques developed in the atomic physics community: optical trapping, transfer to optical tweezers, and full quantum control of individual molecules. Based on the calculations described above, this level of closure is expected to require between 5 and 7 vibrational repumpers. Ref.~\cite{Tarbutt2015a} showed that the forces, and therefore capture velocity, of a typical magneto-optical trap (MOT) grow with increasing excited state $g$-factor; small excited $g$-factors in the $\tilde{A} \, ^2\Pi$ state have been a common problem for molecular MOTs. Excited state $g$-factors of the ATMs discussed here may be large ($\sim$1) due to the quenched orbital angular momentum. This would increase the MOT capture velocity and forces, easing the loading and cooling processes.

\section{Applications} \label{sec:Applications}

Many exciting avenues will be enabled by photon cycling and laser cooling of ATMs. Applications span a broad array of fields, accessing the full breadth of modern physics and chemistry research. In what follows we highlight many promising research directions. It is our hope that this work inspires further exploration of the applications of photon cycling and laser cooling applied to ATMs. 

\subsection{Quantum Computation and Simulation}
\subsubsection{Information Storage}
Ultracold molecules with large electric dipole moments are an exciting new platform for quantum simulation and information processing~\cite{Wall2013,Wall2015,Blackmore2018,Yelin2006,Yu2019}. Recently, Albert, Covey and Preskill~\cite{albert2019robust} have proposed to use 3D rigid rotor molecular codes (i.e. coherent superpositions of several molecular orientations of ATMs) to construct robust encoding of quantum information that is protected against both momentum kicks and orientation diffusion. The structural asymmetry of these three-dimensional rigid rotors, whose orientation is described as an element of the 3-dimensional rotation group SO(3), is exactly what allows for robust encoding of information in the manifold of rotational quantum states. Such quantum error-correcting codes for ATMs like CaSH may enable robust storage and coherent processing of quantum information in a scalable experimentally-accessible manner within the near future. 

The rich internal structure of dipolar molecules also allows the implementation of qudits, higher dimensional analogs of two-level qubits~\cite{Sawant2019}. Compared to qubits, qudits can be more robust~\cite{Zilic2007, Parasa2011} and offer improved error correction~\cite{Campbell2012,Krishna2019}. The number of $d$-level qudits required to produce a Hilbert space of a given size is a factor of $\log_2 d$ smaller than the number of qubits required. Similarly, the computational time to carry out a given set of gate operations using transformations in the $d$-dimensional space is reduced by a factor of $(\log_2 d)^2$~\cite{Sawant2019}. As discussed in Ref.~\cite{Sawant2019}, the rotational and hyperfine levels in dipolar molecules can be used to form a qudit; the authors showed how to construct four-level qudits using diatomic molecules with $^1\Sigma$ and $^2\Sigma$ ground states. The rotational levels in ATMs can be used similarly, although with a much larger number of states available. Importantly, not only are there \textit{more} low-lying (metastable) rotational states to manipulate, but many more of these levels are coupled to one another by dipole allowed transitions~\cite{Patterson2018}. 

\subsubsection{Molecule-Molecule Coupling}
\begin{figure}[t!]
\centering
\includegraphics[width=1\columnwidth]{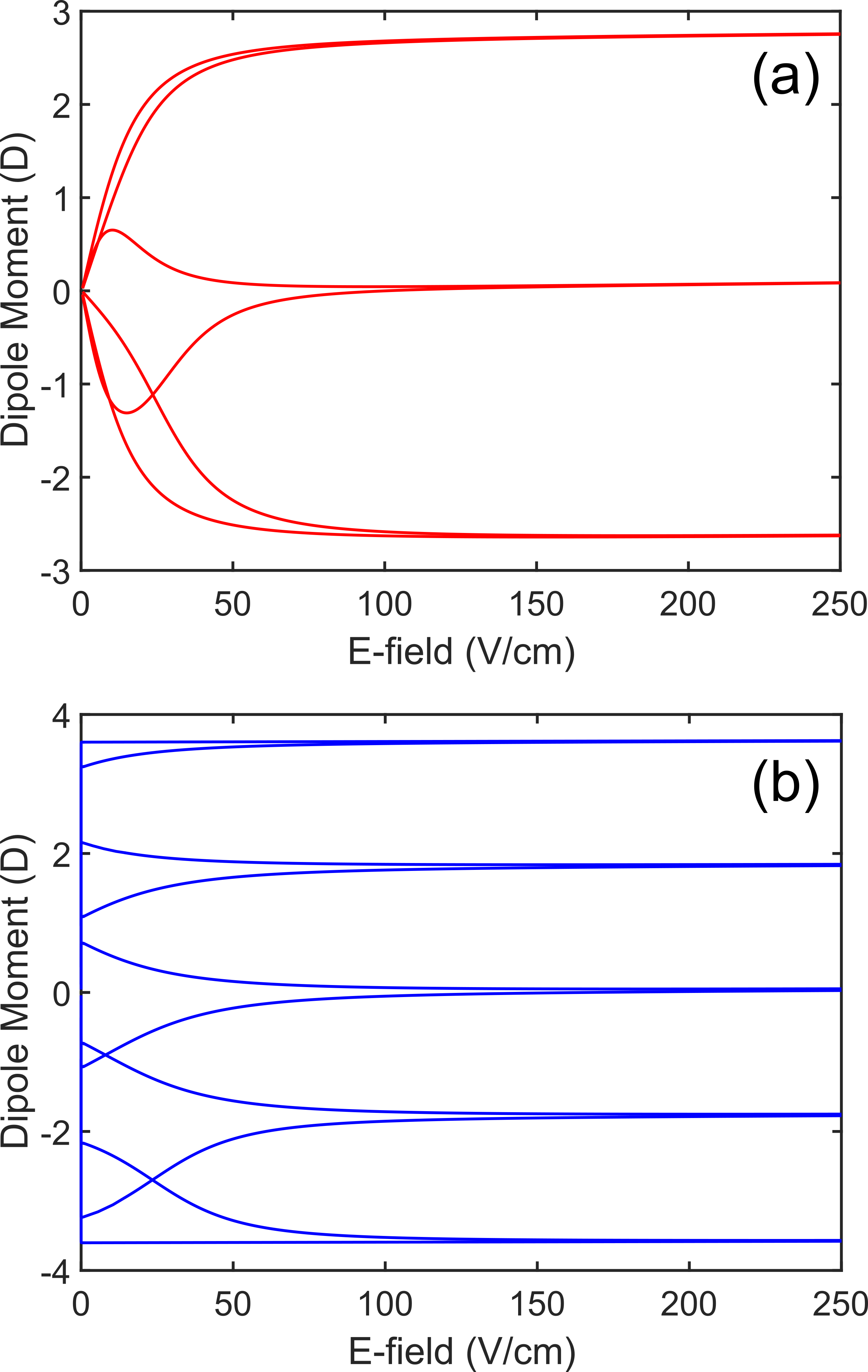}
\caption{Ground state dipole moments of CaSH as a function of electric field. (a) Dipole moments in the $1_{10}/1_{11}$ asymmetry doublet. (b) Dipole moments in the $2_{20}/2_{21}$ asymmetry doublet. Full alignment is achieved at fields $\lesssim 100$~V/cm for $1_{1 K_c}$ and $\ll 100$~V/cm for $2_{2 K_c}$. Note that, even at the maximum fields shown in these plots (250~V/cm), the diatomic species CaF would only achieve a lab-frame dipole moment of 10-100$\times$ smaller.}
\label{fig:StarkAndDipole}
\end{figure}

In addition to quantum information storage, ultracold ATMs are a promising platform for quantum information \textit{processing} due to their strong dipole-dipole interactions. In such experiments, the interaction between two neighboring molecules scales like $d^2$, so choosing species with the largest possible dipole moment is of utmost important. To date, the laser-coolable species proposed for these applications have relatively small dipole moments $\sim 1-2$~D. The diatomic species cooled so far cannot be fully polarized with reasonable laboratory fields, so microwave mixing may be employed which further reduces the achievable dipole moments. Bent ATMs generically have large dipole moments due to the increasingly covalent metal-ligand bond. In addition, they can have multiple dipole moment components, one along each principal axis. For example, CaSH has a large dipole moment of $\sim$5.5~D along the $a$-axis and an additional dipole moment of $\sim$1.5~D along the $b$-axis~\cite{Scurlock1994}.\footnote{The bent structure is an important contributor to this larger $a$-axis dipole moment as can be seen by comparing the cases of CaSH ($d \approx 5.5$~D) and CaOH ($d \approx 1.4$~D)~\cite{Scurlock1994, steimle1992supersonic}.} For the same interparticle spacing, ATMs therefore offer a viable route to achieving molecule-molecule couplings at least an order of magnitude stronger than can be achieved with diatomic species laser cooled to date~\cite{McCarron2018}. Figure~\ref{fig:StarkAndDipole} shows the lab-frame dipole moments of CaSH in several rotational states; note that the molecule is fully polarized at applied fields $\lesssim 100$~V/cm.

As a simple example, consider two fully polarized CaSH molecules in optical tweezers separated by $\sim 1$~$\mu$m. In this case, achievable dipole-dipole interaction rates are $\sim$900~Hz (2~kHz) in the $1_{10}/1_{11}$ ($2_{20}/2_{21}$) manifold. An illustrative comparison can be made against CaF, for which rates $\sim 100$~Hz are expected due to the smaller body-frame dipole moments and the greater difficulty to polarize. For any given separation, we expect the dipolar exchange rates of fully polarized ATMs to exceed their (fully polarized) monofluoride counterpart by about 1-2 orders of magnitude. At the small electric fields shown in Fig.~\ref{fig:StarkAndDipole}, CaSH will be fully polarized while CaF will not be; in this more realistic case, the interaction between two ATMs will exceed that between two CaF molecules by about 3 orders of magnitude.

Use of oriented molecules has been of great interest in recent years~\cite{Hartelt2008,wei2011entanglement,Lin2018}. ATMs with an unpaired electron spin extend these proposals yet further. We propose a simple scheme to polarize the molecule in order to access both of these dipole moments.  A DC electric field $\sim$400~V/cm is used to mix the opposite parity $1_{10}$ and $1_{11}$ states and induce a dipole moment along the $a$-axis. Figure~\ref{fig:StarkAndDipole} shows the Stark structure typical for an MSH molecule in a DC electric field, using the molecular constants of CaSH. The spectrum is essentially identical to that of symmetric tops already proposed for quantum information schemes~\cite{Yu2019}. At the same time, a microwave field is used to mix the Stark-shifted state with one of the $1_{01}$ states. This induces an orientation along the $b$ axis. In this way, the molecule may be aligned along one axis and oriented along an axis perpendicular to it.

Following the very fruitful theoretical work on using diatomic molecules with both electric and magnetic dipole moments for quantum simulation protocols, our hope is that the experimental platform presented here will inspire theoretical inquiry into the use of ATMs with multiple large electric dipole moments.

\subsection{Time-Variation of Fundamental Constants}
Variations of fundamental constants such as the proton-to-electron mass ratio, $\mu=m_p/m_e\sim1836$, are predicted in many models that extend the Standard Model of particle physics~\cite{Safronova2018}. Large-amplitude motions in polyatomic species, such as inversion and hindered rotation, have been predicted to yield orders-of-magnitude enhanced sensitivity to $\mu$ variation~\cite{jansen2014perspective}. These motions are present in any molecule containing an internal rotor with local $C_{3v}$ symmetry, but have no analog in linear or symmetric top molecules. For instance, torsion-rotation transitions in both methanol (HOCH$_3$) and methyl mercaptan (HSCH$_3$) have been found to offer significant enhancement factors~\cite{ kozlov2013microwave, Jansen2013, Jansen2013Methyl}. Table~\ref{tab:MLFCF} includes the laser-coolable analogs of these molecules, e.g. CaSCH$_3$. If lower mass is desired, Be or Mg atoms could be substituted in place of Ca. Similarly, CaOC$_2$H$_5$, CaC$_5$H$_4$CH$_3$, and CaNHCH$_3$ offer methyl groups with hindered internal rotations that may offer large enhancement factors for tests of $\mu$-variation. Efficient state preparation, unit-efficiency readout, phase-space compression for enhanced counting statistics, and long interrogation times enabled by laser cooling may make these systems ideal testbeds.

\subsection{Chiral Species}
The single chirality of many biological molecules (left-handed amino acids but right-handed sugars) has fascinated physicists, chemists, and biologists for over a century. The origin of this biomolecular homochirality remains largely unsolved despite its vital implications for our understanding of the formation and evolution of life on Earth~\cite{Quack2002,Blackmond2011}. Precision spectroscopy of cold chiral molecules, searching for the slight energy difference between enantiomers, could elucidate some aspects of this fundamental symmetry violation. The long interrogation times enabled by traps could greatly enhance the precision of this spectroscopy, but producing trapped samples of chiral methyl derivatives is difficult due to the small electromagnetic moment of these molecules. Direct laser cooling of chiral molecules may represent a viable alternative, one in which photon cycling and laser cooling of an ATM is unavoidable. The crucial advantage of this method is that the laser cooling properties are determined primarily by the  metal atom. Because the parity-violating energy difference scales strongly with nuclear charge, the heavy radicals BaOCHDT or YbOCHDT can be considered without sacrificing good laser cooling properties. Even in the absence of laser cooling to ultralow temperatures, photon cycling of $\sim$100 photons would allow high fidelity quantum state preparation and readout~\cite{lasner2018statistical}.

\subsection{Chalcogens, Thiols, and Hydrocarbons}
Recent work has demonstrated unprecedented control of molecular dissociation, including controlled dissociation in which no additional kinetic energy is imparted to the molecular fragments~\cite{mcguyer2015high}. Such techniques have been proposed as a promising route to produce a variety of ultracold \textit{atoms}~\cite{Lane2012, lane2015production} and can be used to produce molecular fragments such as SH, SCH$_3$, S, NH$_2$, C$_4$H$_4$CH$_3$, or OC$_2$H$_5$ with even lower kinetic energy distributions to their (laser-coolable) parent molecules. 

The chalcogens (S, Se, or Te) display fascinating chemistry which differs dramatically from that of the isoelectronic O atom. This is due to the fact that chalcogens have vacant $d$-orbitals, plus lower electronegativity and higher polarizability as compared to O. The thiols and heavy chalcogens proposed in this work represent an expansion of the chemical diversity of laser-cooled polyatomic species. Thiols and sulfides form self-assembled monolayers far more actively than other functional groups~\cite{Crudden2014}. Recent spectroscopic work has attempted to understand the metal-sulfur bonding by studying the simplest diatomic analog, the AuS molecule~\cite{Kokkin2015}, and lower temperatures may enable more detailed understanding. 

At ultralow temperatures, where targeted chemistry at the bond-specific level is conceivable, one may be able to unravel subtleties of the sulfur-metal bond as compared to its oxygen-metal counterpart. The production methods described in Sec.~\ref{sec:Production} can be readily adapted to produce more exotic chalcogen-containing polyatomic molecules, for example by ablating AE-metals in the presence of selenic acid (H$_2$SeO$_4$) or telluric acid (Te(OH)$_6$).

Metal atoms bound to unsaturated carbon rings have long been of interest to chemistry, dating back to the discovery of the structure of the ``sandwich" molecule ferrocene, Fe(C$_5$H$_5$)$_2$~\cite{Bernath1997}. This led to interest in producing ``half-sandwich" gas-phase molecules like MCp, with Cp = C$_5$H$_5$ the cyclopentadiene ligand. A vibrational analysis for jet-cooled CaCp has been completed~\cite{Robles1992} and rotationally-resolved spectra have also been reported~\cite{Cerny1995}. Monomethyl-substituted analogues, for example CaC$_5$H$_4$CH$_3$ (CaMeCp), have also been observed, as have monopyrrolates, CaC$_4$H$_4$N (CaPy), in which one of the CH groups is replaced by N~\cite{Bopegedera1987alkylamide, Robles1992}. Both CaMeCp and CaPy were found to have the metal atom bonded to the face of the ring. FCFs have been measured for these species, with results that look promising for photon cycling, i.e. diagonal $0_0^0$ FCF $> 0.8$ (see Tab.~\ref{tab:MLFCF}).
Even in these large, complex species, $\sim 100$ photons or more could be cycled with only a few lasers (see Sec.~\ref{sec:Vibrational}). These species represent a versatile playground for organometallic chemistry due to the possibility of synthesizing species with unique substituents around the ring. Laser cooling of AE-bonded hydrocarbon rings, followed by bond-selective dissociation, represents a plausible way to generate samples of isolated, cold carbon rings. This method is particularly well-suited to hydrocarbons which possess neither electric nor magnetic dipole moments (or very small ones) and are therefore difficult to manipulate in the gas phase in their natural form.

\subsection{Astrochemistry}
Many dozens of molecules have now been identified in circumstellar and interstellar environments~\cite{herbst2009,Kwok2016}. These environments are characterized by a wide range of density and radiation conditions which directly influence the abundances and distribution of these molecules. Many of the molecules proposed in this work (e.g., thiols, alcohols, and hydrocarbons) would provide stringent tests of calculated rate constants used in models to describe astrophysical evolution~\cite{Pattillo2018,herbst2009,Kwok2016}. Laser cooling followed by bond-selective dissociation would provide direct access to many of these species, already trapped at cold/ ultracold temperatures. Ultracold molecules will allow long-lived trapping, increasing the measurement precision, and full internal state control, allowing specific states to be populated in order to generate internal rotational/ vibrational temperatures matching the astrophysically relevant scale of 10s to 100s of K. For example, optical cycling will enable the preparation of state distributions for tailored studies of dynamics and rate coefficients necessary to develop comprehensive models of circumstellar and interstellar evolution.~\cite{Faure2011,Smith2011}.

\subsection{Surfaces Functionalized for Optical Cycling}
It has been proposed recently~\cite{Zhao2019} to functionalize surfaces by depositing an optical cycling center on a substrate while sufficiently decoupling the electronic transition from vibrational leakages. This would permit highly scalable quantum devices, e.g. by patterning 2D arrays of cycling centers in close proximity on a substrate or placing these cycling centers in photonic cavities. The strong couplings and individual addressability in such a platform would allow for applications in biophysics, quantum information processing, many-body dynamics, and quantum optics~\cite{Toworfe2009, Lessel2015, Zhao2019}. The ideal ``linker" molecule for attaching optical cycling centers to surfaces is far from certain. Our work shows that AE-nitrogen, AE-sulfur, and AE-carbon ring bonds are favorable for optical cycling, complementing the AE-oxygen bond long recognized for its cycling properties. Self-assembled monolayers with a variety of different surface chemistry have been explored, including OH, NH$_2$, and COOH~\cite{Toworfe2009}. Our present work signals that a wider variety of metal-ligand bonds than previously envisioned may permit photon cycling. Although this research direction remains highly speculative, it can benefit from our suggestion that a more diverse class of AE-ligand bonds can give rise to favorable optical cycling properties.

\section{Conclusion}
In this paper, we analyze how the discrete breakdown of structural symmetry in polyatomic molecules influences repeated optical cycling and direct laser cooling -- the two cornerstones of modern atomic physics and its vast applications. Despite complications due to the reduced symmetry, we show that for a generic class of asymmetric top molecules optical cycling and laser cooling are possible. Our calculations, combining semi-empirical and \textit{ab initio} methods, demonstrate that a wide array of metal-ligand bonds have favorable optical cycling properties, introducing nitrogen- and sulfur-containing asymmetric compounds as laser cooling candidates. We also discuss how some of the approximations used in these calculations may break down, while showing that this should not significantly hamper the optical cycling possibilities. The experimental outlook is quite bright, thanks in part to the ability to produce large numbers of these molecules in cryogenic buffer-gas beam sources. Finally, we outlined many potential applications of optical cycling and laser cooling of asymmetric top molecules; these spanned a broad range of disciplines from chemistry and surface science to quantum simulation and quantum information processing. Our hope is that this work will spur robust theoretical and experimental exploration, similar to early proposals on diatomic molecules with optical cycling centers. Ultracold asymmetric top molecules represent an exciting development at the intersection of physics and chemistry. 

Additional work is necessary to complement the calculations presented here. In particular, improved experimental measurements of the vibrational branching ratios in these species would further benchmark our calculations. Measurements accurate to the $10^{-3}$ level or better will be necessary to assess the viability of laser cooling for any given species. This level of accuracy is possible with current spectroscopic techniques~\cite{Smallman2014}. Improved theoretical calculations can also aid searches for species suited to laser cooling. \textit{Ab initio} methods have previously applied to  develop a rational basis for molecule selection for linear species~\cite{Ivanov2019Rational}. Extensions to non-linear, asymmetric species are now desirable.

\textit{Note} -- During the course of manuscript preparation we became aware of a related and complementary work by K\l{}os and Kotochigova~\cite{Klos2019}, where they used \textit{ab initio} methods to calculate Franck-Condon factors for various polyatomic molecules with increasing complexity.

\section{Acknowledgments}
We gratefully acknowledge Timothy Steimle for his comments on this manuscript and for enthusiastically sharing his knowledge of molecular spectroscopy over many years. We thank Zack Lasner, Louis Baum, Wes Campbell, and Justin Caram for useful discussions. Work at both Harvard and Columbia has been supported by the W. M. Keck Foundation and at Harvard by the Department of Energy. BLA acknowledges support from the NSF GRFP. IK was supported by the Simons Junior Fellow Award.

\appendix
\section{Born-Oppenheimer Approximation Breakdown} 
\label{sec:BOBreakdownApp}
We have generally been working within the Born-Oppenheimer (BO) approximation, in which potential energy surfaces are assumed to be well-separated and non-interacting. In reality, vibronic coupling between these adiabatic surfaces can affect this picture~\cite{Fischer1984}. BO approximation breakdown is especially likely as the size of the molecule grows because the number of states and the probability of near-degeneracies increases. Furthermore, breakdown of the BO approximation is far more likely in the excited electronic states $\tilde{A}$, $\tilde{B}$, and $\tilde{C}$ due to their close proximity.

BO approximation breakdown can alter the intensities of rovibronic decays as compared to those calculated above. This is not necessarily detrimental to photon cycling and laser cooling, so long as the branching induced by vibronic coupling is contained within a small number of modes which are easily repumped. Whether this is the case will require a careful, specialized analysis for any given molecule of interest. Here, we outline general features of such analyses. 

The effects of vibronic coupling can be described by introducing a term, $H_{ev}$, to the Hamiltonian which which mixes two (unperturbed) electronic states due to the influence of vibrations. It is natural to expand this perturbation term in the normal coordinates, yielding to first order~\cite{Henderson1966}

\begin{equation}
    H_{ev}(x,Q) = H_{ev}^0(x,0) + \sum_{j} \left( \frac{\partial H_{ev}}{\partial Q_j} \right)_{0} Q_j,
    \label{eq:BOBreakdownHamiltonian}
\end{equation}

\noindent where $x$ are the electronic coordinates and $Q_j$ the vibrational normal modes. The physical content of Eq.~\ref{eq:BOBreakdownHamiltonian} is that vibronic perturbations can be introduced due to direct coupling between two vibronic states (the 0$^\text{th}$ order term) or by a vibrational ``promoter" mode (the 1$^\text{st}$ order term). 

Because the Hamiltonian belongs to the totally symmetric representation, there are restrictions on which vibronic levels can interact under the influence of Eq.~\ref{eq:BOBreakdownHamiltonian}. For instance, because $H_{ev}^0(x,0)$ is totally symmetric, it can connect the $\tilde{A} \, ^2A^\prime$ and $\tilde{C} \, ^2A^\prime$ states in $C_s$ symmetry. It cannot connect any of the three lowest electronically excited states ($\tilde{A}\, ^2B_2$, $\tilde{B} \, ^2B_1$, and $\tilde{C} \, ^2A_1$) in $C_{2v}$ symmetry because all three of these states have different symmetry. 

The 1$^\text{st}$ order term in Eq.~\ref{eq:BOBreakdownHamiltonian} is a linear sum over the normal modes, so we consider each term in the sum individually. The product $\left( \frac{\partial H_{ev}}{\partial Q_j} \right)_{0} Q_j$ must be totally symmetry, so the symmetry of $\left( \frac{\partial H_{ev}}{\partial Q_j} \right)_{0}$ can be determined once the normal mode symmetries are known. As an example, in $C_{2v}$ a vibrational mode with symmetry $b_2$ can induce coupling between vibronic states with symmetries $A_2$ and $B_1$. 

While these considerations can help predict the \textit{propensity} for BO approximation breakdown, they do not provide any estimate of the beakdown's \textit{magnitude}. A simple estimate for the impact of BO approximation breakdown can come from the case of linear molecules which have been previously laser cooled. Typical perturbation matrix elements are in the range of $\sim 0.5-5$~cm$^{-1}$ for CaOH or SrOH when Renner-Teller vibronic interactions are considered~\cite{Presunka1994,Li1992}. Taking this scale as the typical matrix element, the strength of vibronic interactions can be estimated using second order perturbation theory and typical energy splittings in the excited electronic states. Such splittings are typically $\sim 100-1000$~cm$^{-1}$, implying that BO breakdown can induce unexpected decays at the $10^{-3}-10^{-6}$ level. In the case of very near degeneracies or enhanced coupling matrix elements this interaction can be much stronger.

\section{Details of \textit{ab initio} calculations} \label{sec:AbInitioApp}

\textit{Ab initio} calculations were performed using the ORCA quantum chemistry program~\cite{neese2012orca}. Molecular orbitals, optimized geometries, normal modes and vibrational frequencies for all the ground and excited states were calculated at the level of unrestricted Kohn-Sham (UKS) DFT using the B3LYP functional in the ORCA package with def2-TZVPP basis following other similar works~\cite{ORourke2019, isaev2015polyatomic}. The calculations for the excited states used TDDFT. Table~\ref{tab:AbInitioComparisons} provides a comparison of our calculations for the lowest three electronic excitations with the experimentally measured values. Our agreement is comparable with other reported \textit{ab initio} calculations for such species~\cite{ortiz1990CaSH,koput2002CaOH,taylor2005electronic}. Tables \ref{tab:CaSHgeom}, \ref{tab:CaSH_TDMs}, \ref{tab:CaNH2geom}, and \ref{tab:CaNH2_TDMs} provide optimized geometries for the first excited state as well as predicted transition dipole moments for CaSH and CaNH$_2$. 

\begin{table}[b]
\begin{tabular*}{\columnwidth}{@{}c@{\extracolsep{\fill}}cccccc@{}}
\hline \hline Molecule & Calc. 1 & Calc. 2 & Calc. 3 & Meas. 1 & Meas. 2 & Meas. 3\tabularnewline
\hline 

CaOH & 16699 & 16699 & 18363 & 15998 & 15998 & 18022\tabularnewline
\hline 
CaSH & 14812 & 15212 & 15431 & 15381 & 15859 & 16075\tabularnewline
\hline 
CaNH$_2$ & 16144 & 16406 & 17605 & 15464 & 15885 & 17375\tabularnewline \hline \hline 
\end{tabular*}

\caption{Benchmarking comparison of the ab initio predicted and experimentally
measured 3 lowest electronic transition energies. Values are provided in cm$^{-1}$. Experimental measurements are obtained from Refs.~\cite{bernath1985CaOHX2B, dick2006CaOH, Sheridan2005SrNH2}}. 

\label{tab:AbInitioComparisons}
\end{table}

\begin{table}
\begin{centering}
\begin{tabular*}{0.75\columnwidth}{@{}c@{\extracolsep{\fill}}ccc@{}}
 &  &  & \tabularnewline
\hline 
\hline 
Atom & X (\AA) & Y (\AA) & Z (\AA)\tabularnewline
\hline 
Ca & 0 & -0.045481 & -0.062492\tabularnewline
S & 0 & -0.037570 & 2.538967\tabularnewline
H & 0 & 1.364911 & 2.651461\tabularnewline
\hline 
\hline 
 &  &  & \tabularnewline
\end{tabular*}
\par\end{centering}
\caption{Optimized geometry for CaSH in the first exited electronic state.
Cartesian coordinates are in Angstroms. }
\label{tab:CaSHgeom}
\end{table}

\begin{table}
\begin{centering}
\begin{tabular*}{0.75\columnwidth}{@{}c@{\extracolsep{\fill}}ccc@{}}
 &  &  & \tabularnewline
\hline 
\hline 
Excited State & Tx (au) & Ty (au) & Tz (au)\tabularnewline
\hline 
1 & 0.00000 & 2.25849 & -0.02413\tabularnewline
2 & 2.17414 & 0.00000 & 0.00000\tabularnewline
3 & 0.00000 & 0.00540 & 1.73561\tabularnewline
\hline 
\hline 
 &  &  & \tabularnewline
\end{tabular*}
\par\end{centering}
\caption{Calculated transition electric dipole moments in atomic units for
CaSH for different electronic states.}
\label{tab:CaSH_TDMs}
\end{table}

\begin{table}
\begin{centering}
\begin{tabular*}{0.75\columnwidth}{@{}c@{\extracolsep{\fill}}ccc@{}}
 &  &  & \tabularnewline
\hline 
\hline 
Atom & X(\AA) & Y(\AA) & Z(\AA)\tabularnewline
\hline 
Ca & 0 & 0 & -0.013610\tabularnewline
N & 0 & 0 & 2.102409\tabularnewline
H & 0 & 0.806437 & 2.718700\tabularnewline
H & 0 & -0.806437 & 2.718700\tabularnewline
\hline
\hline 
&  &  & \tabularnewline
\end{tabular*}
\par\end{centering}
\caption{Optimized geometry for CaNH$_{2}$ in the first exited electronic
state. Cartesian coordinates are in Angstroms.}
\label{tab:CaNH2geom}
\end{table}

\begin{table}
\begin{centering}
\begin{tabular*}{0.75\columnwidth}{@{}c@{\extracolsep{\fill}}ccc@{}}
 &  &  & \tabularnewline
\hline 
\hline 
Excited State & Tx (au) & Ty (au) & Tz (au)\tabularnewline
\hline 
1 & 0.00000 & -2.30296 & 0.00000\tabularnewline
2 & 2.11105 & 0.00000 & 0.00000\tabularnewline
3 & 0.00000 & 0.00000 & -1.54932\tabularnewline
\hline 
\hline 
 &  &  & \tabularnewline
\end{tabular*}
\par\end{centering}
\caption{Calculated transition electric dipole moments in atomic units for
CaNH$_{2}$ for different electronic states.}
\label{tab:CaNH2_TDMs}
\end{table}

\section{Rotational Structure}
\label{sec:RotationalApp}

The rotational Hamiltonian for an ATM in the principal axis system, including the spin-orbit interaction, can be expressed as~\cite{Hirota1985, Morbi1997}

\begin{equation}
H = A (N_a - L_a)^2 + B (N_b-L_b)^2 + C (N_c-L_c)^2 + \sum_{j} \xi \mathbf{\ell_j} \cdot \mathbf{s_j},
\label{eq:RotHam}
\end{equation}

\noindent where $\hat{N}$ is the total angular momentum excluding spin, $\hat{L}$ is the electronic orbital angular momentum, $\xi$ characterizes the spin-orbit interaction, and the sum over $j$ includes all electrons in the molecule. Following Refs.~\cite{Shirley1990, Morbi1997}, Eq.~\ref{eq:RotHam} can be reduced to a purely rotational Hamiltonian,

\begin{equation}
H_\text{rot} = A N_a^2 + B N_b^2 + C N_c^2,
\label{eq:PureRotHam}
\end{equation}

\noindent with perturbations arising from cross terms 

\begin{equation}
\begin{split}
H^\prime = -2 & \left( A N_a L_a + B N_b L_b + C N_c L_c \right ) \\
 &+ \sum_{j} \xi \mathbf{\ell_j} \cdot \mathbf{s_j}.
\end{split}
\label{eq:PerturbationTerms}
\end{equation}

\noindent A Van Vleck transformation produces an effective spin-rotation Hamiltonian~\cite{Morbi1997,Shirley1990, Hirota1985}

\begin{equation}
H_\text{sr} = \frac{1}{2} \sum_{\alpha,\beta} \epsilon_{\alpha \beta} \left( N_\alpha S_\beta + S_\beta N_\alpha \right),
\label{eq:SRHam}
\end{equation}

\noindent that operates only within a given electronic state. The elements of the spin-rotation tensor, $\epsilon_{\alpha \beta}$, contain first-order contributions due to direct coupling between electron spin and molecular rotation and second-order contributions due to cross terms with the spin-orbit interaction. These second order contributions are often a large portion of the $\epsilon_{\alpha \beta}$ components. While $\epsilon_{\alpha \beta}$ has 9 possible parameters, for molecules with orthorhombic symmetry only the three parameters $\epsilon_{aa}$, $\epsilon_{bb}$, and $\epsilon_{cc}$ are nonzero. The rotational constants are also contaminated by second order contributions, although these are generally smaller in magnitude than the first-order (rigid body rotation) terms. We emphasize that careful deperturbation of the constants is important to assign the accurate geometries necessary for FCF calculations.

For the molecules considered in this paper the hyperfine structure is negligible when the only nuclei with nonzero spin are located multiple bond lengths away from the valence electron. (To date, only spin-0 Ca and Sr nuclei have been used in molecular laser cooling experiments.) If fermionic  metals are considered as the optical cycling center, these hyperfine terms must be added and the analysis proceeds as a straightforward extension from that presented in this work. Hyperfine interactions can mix rotational states with selection rules $\Delta N = \pm 2$, $\Delta F = 0$~\cite{BrownCarrington, Lu2014}. However, these couplings are very weak~\cite{Lu2014} and have not affected laser cooling of molecules to date.

To compute energies and decay strengths, we construct a numerical representation of the Hamiltonian, $H = H_\text{rot} + H_\text{sr}$, in a Hund's case (b) symmetric top basis, $\lvert N, K, S, J, M_J \rangle$, using matrix elements taken from the literature~\cite{Hirota1985,Bowater1973}. Numerical diagonalization yields eigenenergies and eigenstates with a block structure as expected on the basis of symmetry consideration~\cite{Gordy1984}. When $K_a$ ($K_c$) sets the dominant level structure, the $K_c$ ($K_a$) label distinguishes the two parity components with a given $J$. 

Note that $H_\text{rot}$ contains matrix elements with $\Delta K = \pm 2$ selection rules, although these are still diagonal in $N$ and may only affect the low-$N$ states of interest to laser cooling through higher order perturbations. $H_\text{sr}$ includes both diagonal ($\langle N K \rvert H_\text{sr} \lvert N K \rangle$) and off-diagonal ($\langle N K \rvert H_\text{sr} \lvert N K \pm 2 \rangle$, $\langle N K \rvert H_\text{sr} \lvert N-1, K \rangle$, $\langle N K \rvert H_\text{sr} \lvert N-1, K\pm 2 \rangle$, and $\langle N K \rvert H_\text{sr} \lvert N, K\pm 1 \rangle$) elements. The matrix elements that couple states with $\Delta K = \pm 2$ scale with $\left( \epsilon_{bb}^{\prime \prime} - \epsilon_{cc}^{\prime \prime} \right)$ and those with $\Delta K = \pm 1 $ (which vanish for molecules with orthorhombic symmetry) scale with $\left( \epsilon_{ab}+\epsilon_{ba} \right)$~\cite{Hirota1985}. As discussed in App.~\ref{sec:RotationalBranchingApp} both of these are expected to be negligible. The matrix elements that preserve $K$ but couple rotational states with $\Delta N = \pm 1$ scale directly with $\epsilon_{aa}$ and $\frac{1}{2} \left( \epsilon_{bb}+\epsilon_{cc}\right)$. As will be discussed in App.~\ref{sec:RotationalBranchingApp}, these terms are not entirely negligible but likely affect photon scattering only after $\gtrsim 10^5$ optical cycles.

The eigenstates are most conveniently described by the labels $N_{K_a K_c}$, where $N$ is the rotational quantum number, and $K_a$ and $K_c$ are labels which correlate to the projection of $N$ onto the $a$ and $c$ axes in the prolate and oblate symmetric top limits, respectively. When we adopted the example of near-prolate ATMs in Fig.~\ref{fig:RotationalRepumping}, the small moment of inertia along the $a$-axis means that levels of different $K_a$ values are widely spaced, each containing a ladder of $N$ rotational states. For each $N$, there are sublevels split by the spin-rotation interaction as the asymmetry doubling. In general, the spin-rotation splitting grows with increasing $N$ while for a given $N$ the asymmetry doubling shrinks rapidly with increasing $K_a$. It is straightforward to construct a similar diagram for ATMs with any other value of the Ray's asymmetry parameter $\kappa\equiv\left(2B-A-C\right)/\left(A-C\right)$ although the relative energies will shift. For near-oblate species, the roles of $K_a$ and $K_c$ are swapped, but otherwise the conclusions remain the same.

\section{Rotational Branches and Closure} 
\label{sec:RotationalBranchingApp}

Electric dipole selection rules can be determined by symmetry considerations applied to the rovibronic states. In particular, electric dipole transitions must obey the constraint~\cite{BunkerJensen}

\begin{equation}
    \Gamma^{\prime \prime} \otimes \Gamma^{\prime} \supset \Gamma^{\ast},
\end{equation}

\noindent where $\Gamma^{\prime \prime}$ and $\Gamma^{\prime}$ are the symmetries of the rovibronic states involved in the transition, and $\Gamma^{\ast}$ is the electric dipole representation of the molecular symmetry group. General selection rules on the values of $K_a$ and $K_c$ can be derived by working in the rotation group $D_2$. 
It is useful to compare the rotational selection rules (see Tab.~\ref{tab:RotSelectionRules}) to the case of symmetric top molecules. Consider first a prolate symmetric top molecule, e.g. CaCH$_3$ (see Fig. \ref{fig:MolOrbs} column 2), in which the $a$-axis corresponds to the symmetry axis, and the $b$- and $c$-axes lie in a plane perpendicular to this. In this case, transitions can be characterized as ``parallel" or ``perpendicular" depending on whether the transition dipole moment is along the symmetry axis or perpendicular to it. Parallel transitions have the selection rule $\Delta K = 0$ while perpendicular transitions have the selection rule $\Delta K = \pm 1$.  As the molecule distorts from symmetric, the equivalence between the $b$- and $c$-axes is broken. Nonetheless, for a near-prolate symmetric top the $a$-type transition is still of approximate ``parallel" nature, and the $a$-type selection rules in Tab.~\ref{tab:RotSelectionRules} are justified. Similar reasoning provides an intuitive basis for selection rules in approximately ``perpendicular" transitions, as well as for near-oblate tops.

ATMs with $C_{2v}$ symmetry are expected to follow these patterns rigorously, as the three components of $\hat{\mu}$ belong to a different irreducible representation of the symmetry group~\cite{BunkerJensen}. This is also reflected in the alignment of the principal axes with the electronic orbitals. In the more general case of $C_{s}$ symmetry, a given excited state can decay by coupling to one of several components of $\hat{\mu}$; for example, both $\hat{\mu_a}$ and $\hat{\mu_b}$ belong to $A^\prime$. Geometrically, the transition moments, determined by the electronic wavefunctions, and the principal axes, determined by the geometry of the molecule, are misaligned. This means that either $K_a$ or $K_c$, or both, can change in a generic electronic decay. For example, in CaSH the $\tilde{C} \rightarrow \tilde{X}$ transition can couple to either $\hat{\mu}_a$ or $\hat{\mu}_c$. In addition, off-diagonal couplings between electronic manifolds due to a term $((\epsilon_{ab}+\epsilon_{ba})/2)$ in the spin-rotation Hamiltonian can mix the rotational state labels $K_a$ and $K_c$. The magnitude of this effect will vary from species to species depending on the spin-rotation tensor.

A simple estimate for the relative intensity of $c$-type transitions to that of $a$-type transitions in molecules with $C_s$ symmetry can be obtained by considering the relative alignment between the principal axes (which determine the rotational structure) and the metal-ligand bond (which determines the local symmetry of the valence orbitals). In the case of molecules like CaSH and SrSH, the angle between the metal-sulfur bond and the $a$-axis is typically $\sim 0.5^{\circ}$. This would imply that roughly 1 in $10^4$ decays from $\tilde{C} \rightarrow \tilde{X}$ couples to $\hat{\mu}_c$. This number is consistent with the transition dipole moments computed by our \textit{ab initio} calculations in Tab.~\ref{tab:CaSH_TDMs}, which predict transition strengths that differ by a factor of $\sim 100$. This means that, if fewer than $10^4$ photon scatters are required, laser cooling of CaSH can proceed using only the scheme from a single panel of Fig.~\ref{fig:RotationalRepumping}. To scatter additional photons without population loss, either microwave remixing or additional laser frequency components can be used to repump population which decays due to these small branches. 

The effect of state mixing, Eq.~\ref{eq:PerturbationTerms}, is most prominent in the electronically excited states. For example, rotation about the $a$-axis can induce mixing between the $\tilde{A}$ and $\tilde{B}$ states which will tend to unquench the orbital angular momentum. Ref.~\cite{Whitham1990} provides a physical picture for these interactions: in the non-rotating molecule, the $\tilde{A}$ and $\tilde{B}$ states correspond to standing $p \pi$ orbitals--- one along the $b$-axis, and one along the $c$-axis--- with quenched orbital angular momentum. As the molecule begins to rotate, the $p \pi$ orbitals follow adiabatically, but eventually decouple from the nuclear framework as a net orbital angular momentum $\langle L_z \rangle$ is induced. This is reflected, for example, in the values of $\epsilon_{aa}$ measured for the hydrosulfides and monoamides~\cite{Whitham1990, Brazier2000}. The magnitude of this interaction grows with increasing $K_a$ and in the limit of large $K_a$, $\epsilon_{aa} \rightarrow A_\text{SO}$. Similarly, rotation about the $b$-axis can lead to interactions between $\tilde{A}$ and $\tilde{C}$ states, and rotation about the $c$-axis can mix the $\tilde{B}$ and $\tilde{C}$ states~\cite{Morbi1997}. 

Whether, and to what degree, these interactions affect laser cooling will differ from one species to another. Generally speaking, for near-prolate species the interaction between $\tilde{A}$ and $\tilde{B}$ states is of little concern, as rotationally closed optical cycles with hybrid $bc$-type selection rules differ only by which subset of asymmetry doublet components are driven by the laser (see Fig.~\ref{fig:RotationalRepumping}(b) and (c)). The full set of levels typically covers $\ll 1$~GHz and can easily be addressed with a single laser. Mixing between the $\tilde{C}$ state and either of $\tilde{A}$ or $\tilde{B}$ introduces bands with selection rule $\Delta K_a = 0, \pm 1$, requiring the addition of one additional laser per vibrational band. Alternatively, microwave remixing can be used to return population to the optical cycle. This is very similar to the case of symmetric top molecuels~\cite{kozyryev2016MOR}. For near-oblate molecules the same ideas apply, though with the role of the $a$- and $c$-axes reversed. Finally, as the orbital angular momentum of the $\tilde{A}$ or $\tilde{C}$ state is unquenched ($\langle L_z \rangle \rightarrow 1$) the level structure will approach that of a $^2 \Pi$ state in symmetric species. In such a case, an additional rotational decay to the $2_{11}, J^{\prime \prime}=3/2$ level may increase in strength. This is analogous to the decay to the $N=2, K=1, J=3/2$ parity doublet in symmetric top species. As was shown in Ref.~\cite{kozyryev2016MOR}, this level is straightforward to remix into the optical cycle. 

Interactions among the electronically excited states may unquench the electronic orbital angular momentum and induce transitions with $\Delta N = \pm 2$. The relative strength of such decays will vary from species to species, and for excited states with spin-rotation constants $\epsilon_{aa}$ small relative to the spacing between vibronic states levels (as is common) the probability of these decays will approach zero. For instance, $\epsilon_{aa}^\prime / \Delta E_{\tilde{B}-\tilde{A}} \sim 10^{-2}$ in the molecules CaSH and SrSH~\cite{Sheridan2007}. Thus this mixing will likely be relevant once more than $\sim 3 \times 10^4$ photons are scattered. Note, too, that the $\Delta N = \pm 2$ rovibronic decay will be induced as the $\tilde{A}$ and $\tilde{B}$ states coalesce into the $\tilde{A} \, ^2E$ state in molecules with near-$C_{3v}$ symmetry. This requires the addition of a laser frequency (or microwave) to repump the $2_{1 K_c}, J^{\prime \prime}=3/2$ state, as discussed in Ref.~\cite{kozyryev2016MOR}. Note that this is the same state requiring remixing for reasons already discussed above, i.e. this is not a \textit{new} loss channel.

Off-diagonal terms of the spin-rotation interaction within the ground state may also be relevant. $H_\text{sr}$, may mix states that differ by $\Delta K_a = \pm 2$ and by $\Delta N = \pm 1$. For the ATMs proposed here, the $\Delta K = \pm 2$ terms are proportional to $\left( \epsilon_{bb}^{\prime \prime} - \epsilon_{cc}^{\prime \prime} \right) \approx 10^{-4}$~cm$^{-1}$~\cite{Sheridan2007}. Compared to the typical spacing between states differing in $K_a$ by $\pm 2$ ($\approx 10-100$~cm$^{-1}$), such terms lead to rotational losses at the $\lesssim 10^{-8}$ level and are negligible. Similarly, for non-orthorhombic molecules there will be matrix elements of $H_\text{sr}$ that can mix states with $\Delta K = \pm 1$. Such matrix elements are proportional to $\left( \epsilon_{ab} + \epsilon_{ba} \right)/2$~\cite{Hirota1985}, which is typically small ($\sim 10^{-4}$~cm$^{-1}$ in CaSH~\cite{Sheridan2007}) compared to the spacing between connected levels. Again, we expect this to lead to mixings at the $\sim 10^{-8}$ level, which is largely negligible. The $\Delta N = \pm 1$ terms in $H_\text{sr}$ are potentially more relevant. Mixing of the $2_{K_a K_c}$ rotational level into the $1_{K_a K_c}$ ground state is most likely to affect the rotationally closed optical cycles. At first order in perturbation theory, the admixed amplitude will be $\zeta \sim \epsilon_{\alpha \alpha}/4 B_\alpha$, where $\alpha = a, b, c$ labels the principal axes. Typical values of $\zeta \approx 1-5 \times 10^{-3}$~\cite{Brewster2000,Janczyk2003}. Such mixing will present a relevant loss channel after scattering $\gtrsim 5 \times 10^4$ photons, more than required for typical laser cooling. If such losses begin to affect the optical cycle, remixing via a single additional laser frequency or microwave transitions can be used to increase the photon budget. Once again, this is the same $N^{\prime \prime}=2$ state that may require repumping for other reasons, so these laser frequencies may already be present.


\bibliography{ATMCooling_library}

\end{document}